\documentclass{aa}
\usepackage{graphics}

\begin{document}

\thesaurus{08:10.07.2;10.15.1;09.04.1}

\title{Foreground and background dust in star cluster directions}

\author{C.M. Dutra \inst{1} \and  E. Bica \inst{1} }

\offprints{C.M. Dutra - dutra@if.ufrgs.br}
 
\institute{Universidade Federal do Rio Grande do Sul, IF, 
CP\,15051, Porto Alegre 91501--970, RS, Brazil}

\date{Received; accepted}

\maketitle

\titlerunning{Reddening in the direction of star clusters}
\authorrunning{C. M. Dutra \& E. Bica}

\begin{abstract}

This paper compares reddening values E(B-V) derived from the stellar content of 103 old open clusters and 147 globular clusters of the
Milky Way  with those derived from DIRBE/IRAS 100 $\mu$m dust emission in the same directions. Star clusters at $|{\it b}|> 20^{\circ}$ show comparable reddening values between the two methods, in agreement with the fact that most of them are located beyond the disk dust layer.
For very low galactic latitude lines of sight, differences occur
in the sense that DIRBE/IRAS reddening values can be substantially larger, suggesting effects due to the depth distribution of the dust. 
The differences appear to arise from dust in the background of the clusters consistent with a dust layer where important extinction occurs up to distances from the Plane of  $\approx$ 300 pc. For 3\,\% of the sample a significant background dust contribution might be explained  by higher  dust clouds. We find evidence that the Milky Way dust lane and higher dust clouds are  similar to those of several edge-on spiral galaxies recently studied in detail by means of CCD imaging.


\end{abstract}

\section{Introduction}

Full-sky  surveys in the far infrared have been achieved
by means of the IRAS and COBE satellite observations. Schlegel et al. (1998) built a reddening map 
from the  100 $\mu$m IRAS dust emission distribution considering temperature
 effects using 100/240 $\mu$m DIRBE data. The transformation to E(B-V) maps was 
 obtained from dust columns calibrated via the (B-V)-Mg2 relation for early type galaxies.
  This far-infrared reddening 
(hereafter E(B-V)$_{\rm FIR}$) presents a good agreement at high galactic latitudes
 with that derived from H I and galaxy counts by Burstein \& Heiles 
 (1978, 1982) with an offset of 0.02 mag (lower values for the  latter method).
 Recently, Hudson (1999)  analysed  E(B-V)$_{\rm FIR}$  maps using
 50 globular clusters with $|b| > 10^{\circ}$ and distance from the plane $|Z| > 3$ kpc, as well as
  86 RR Lyrae from Burstein \& Heiles (1978). These two samples provided slightly
  lower values on the average as compared to Schlegel et al.'s reddening values ($\Delta$E(B-V) = -0.008 and -0.016,
   respectively). The reddening comparisons above hardly exceed the limit E(B-V) $\approx$ 0.30, so that a more extended range should be explored.

   Since the Galaxy is essentially transparent at 100 $\mu$m, the far-infrared
    reddening values should represent dust columns integrated throughout the whole 
    Galaxy in a given direction. Star clusters probing distances as far as
     possible throughout the Galaxy should be useful to study the dust distribution 
     in a given line of sight. Globular clusters and old open clusters are ideal
      objects for such purposes  because they are  in general distant enough to provide a significant probe the galactic interstellar medium and have
       a suitable sky coverage. Clearly, star clusters beyond the disk dust 
       layer are expected to have reddening values essentially comparable to those of galaxies
        in the same direction. On the other hand, clusters within the dust layer
         should have contributions from clouds in background regions.  Another issue is the thickness of the Milky Way dust lane and whether some dust clouds occur at higher distances from the Plane. Recently, several edge-on spiral galaxies have been studied in detail (Howk \& Savage 1999) and  a comparison of their dust distribution with that of the Milky Way is worthwhile.

The aim of the present study is to compare star cluster reddening values measured from direct
methods, i.e. sampling the dust effects seen in  the light emitted by the cluster members, with those derived from the 100 $\mu$m dust emission. We investigate the possibility of background and foreground dust contributions in star clusters directions. In  Sect.2 we present an overview of Schlegel et al.'s (1998) reddening values predicted in different environments in the Galaxy. In Sect.3 we gather the necessary data for globular clusters and old open clusters and describe the sample properties. In Sect.4 we discuss the results, especially the star cluster lines of sight with evidence of background dust. Finally, the concluding remarks are given in Sect.5.     

\section{Overview of dust emission reddening values E(B-V)$_{\rm FIR}$}

For a  better understanding of the reddening distribution throughout the Galaxy
  we extracted E(B-V)$_{\rm FIR}$ values from Schlegel et al.'s  maps using the software {\em dust-getval} provided by them. 
  We discuss (i) directions along the galactic plane which accumulate reddening from sources
   in different arms and  other large structures, and (ii) galactic latitude profiles to
   see the effects of relatively isolated nearby (high latitude) dust clouds.

  We show in Fig.1 the entire Galaxy longitude profile. 
  The upper panel is in direction of the galactic centre and the lower one is in direction of the 
  anticentre. Note the enormous reddening differences between the two panels: the lower panel
   has typical values of E(B-V)$_{\rm FIR}\approx$1.5, and the values in the upper panel are a factor $\approx$ 10 higher. We indicate a series of H\,I, CO and optical features which help interpret the reddening distribution: (i) tangent regions of the spiral arms
     Sagittarius-Carina, Scutum (5 kpc arm) and 4 kpc arm (Henderson 1977, 
     Georgelin \& Georgelin 1970a, Cohen et al. 1980); (ii) the extent of the 3
      kpc arm (Kerr \& Hindman 1970, Bania 1980); (iii) the extent of the far side of
       the Sagittarius-Carina arm (Grabelsky et al. 1988); (iv) the Molecular
        Ring (MR) and the Central Molecular Zone (CMZ), (Combes 1991, Morris \& 
        Serabyn 1996); and finally, (v) the extent of the Local (Orion) and Perseus
         arms (Georgelin \& Georgelin 1970b).

The relatively low reddening in the anticentre panel can be basically explained
 by the cumulative effect of the three external arms: Orion, Perseus and Outer
  arm (Digel et al. 1990). It is worth noting that E(B-V)$_{\rm FIR}$ on the average
   is higher in the second quadrant than in the third quadrant, probably by the 
   interruption of the Perseus arm. The steady increase of E(B-V)$_{\rm FIR}$ in 
   the first and fourth quadrants towards the direction of the Galactic center
    can be explained by the cumulative effect of inner arms and especially their
     tangent zones. Owing to the 100 $\mu$m dust emission transparency the far
      side arms of the Galaxy will also contribute to E(B-V)$_{\rm FIR}$ (see the
       extent of far side of the Sagittarius-Carina arm in the fourth quadrant). The Molecular Ring is also a major contributor, leading to a plateau level E(B-V)$_{\rm FIR}\approx$ 20. Finally, the Central Molecular Zone is responsible for the central cusp.

\begin{figure*} 
\resizebox{\hsize}{!}{\includegraphics{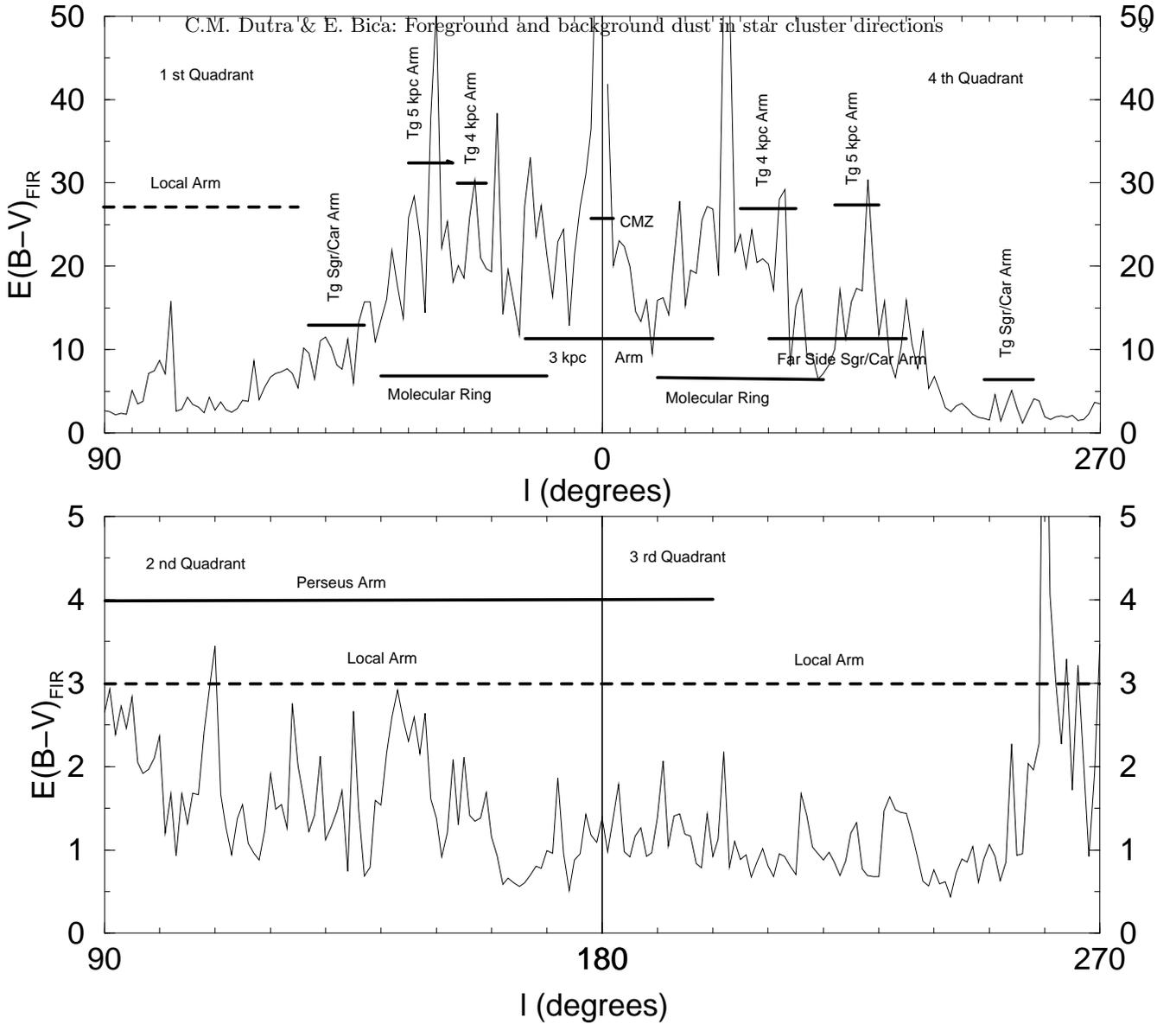}}
\caption[]{Reddening distribution in the Galactic Plane derived from dust emission (Schlegel et al. 1998). The lines indicate the limits in galactic longitude of the spiral arm features and other large scale structures (see the text)}
\label{fig1}
\end{figure*}

Figure 2 shows E(B-V)$_{\rm FIR}$ profiles in the interval $ -25^{\circ} \le {\it b} \le 25^{\circ}$
 for selected galactic longitudes including well-known dark cloud centers.
 The individual clouds, especially their central parts can attain comparable (in
 some cases higher)
  E(B-V)$_{\rm FIR}$ values to disk zones at lower latitudes. Individual dark clouds
  have a core-halo structure. In the $\rho$ Oph dark cloud the core FWHM is 35'
  while at E(B-V)$_{\rm FIR}$ = 0.5 the halo diameter is 4$^{\circ}$. For the
  Chamaleon I complex the core FWHM is 48' (in the region of the reflection nebula IC\,2631)  
  while the halo diameter at E(B-V)$_{\rm FIR}$ = 0.5  is 2.9$^{\circ}$. 
  From these simple comparisons and corresponding solid angles the zones 
  responsible for the accumulation of reddening throughout the arms
  appear to be the cloud halos rather than the cores, possibly combined with diffuse
   galactic dust.

\begin{figure*} 
\resizebox{\hsize}{!}{\includegraphics{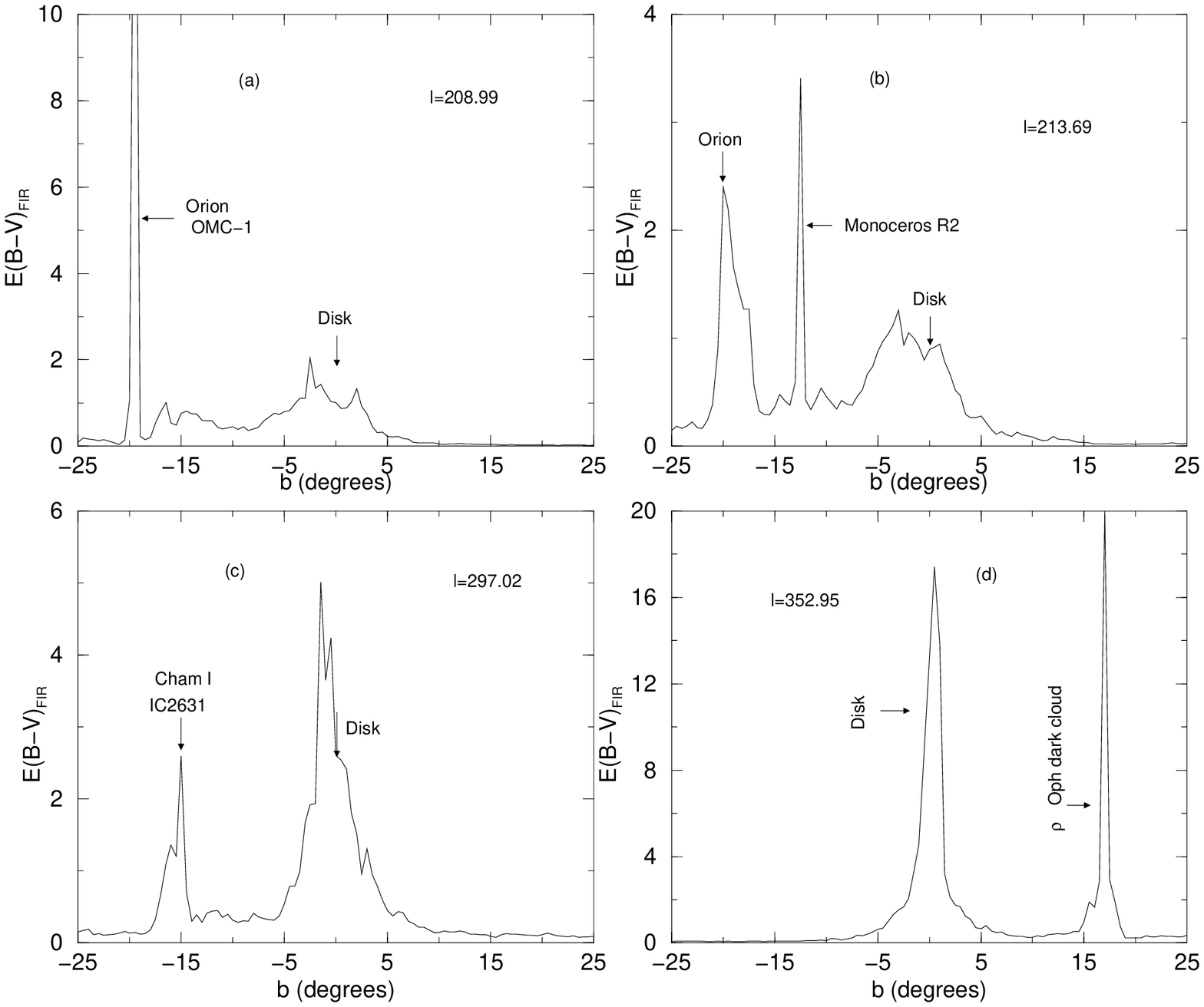}}
\caption[]{E(B-V)$_{\rm FIR}$ reddening profiles in galactic latitude including relatively high
 dark clouds: (a) Orion Molecular Cloud 1, (b) Monoceros R2, (c) Chamaleon I dark cloud centred
  at the reflection nebula IC\,2631, and (d) $\rho$ Ophiuchi dark cloud centred at 
  the molecular cloud core IR star cluster. The longitude is indicated in 
  each panel}
\label{fig2}
\end{figure*}

\subsection{Comparison with reddening values from JHK photometry for nearby dark clouds}

We show in Table 1 embedded infrared star clusters and T Tauri groups which are
    related to nearby dust complexes. They can be useful to analyse the high
    extinction regime and to compare Schlegel et al.'s reddening values with those measured
    directly from Infrared (JHK) photometry of the stellar content. We give 
    E(B-V)$_{\rm FIR}$ values for the central positions of these stellar
     aggregates and indicate to which complexes they belong. The identifications
      and locations of these objects are from: (i) Gomez \& Lada (1998) for the
       T Tauri groups related to the dark clouds Barnard 30 and 35; 
       (ii) Lada et al. (1991) for the infrared clusters embedded in the
        nebulae NGC\,2071, M\,78, NGC\,2023 and NGC\,2024
       in the Orion Complex LDN\,1630 Molecular Cloud; (iii) Minchin et al. (1991)
       , Jones et al. (1994) and Reipurth et al. (1999 and references therein)
        respectively for the three deeply embedded
       clusters OMC-1, OMC-2 and OMC-3 in Orion Complex Molecular clouds;
       (iv) Strom et al. (1993) 
      for the 7 objects in the  Orion complex LDN\,1641 Molecular Cloud; 
      (v) Carpenter et al. (1997)
       for the Mon R2 IR cluster; (vi) Lawson et al. (1996) for two 
       concentrations of T Tauri stars around the reflection nebulae IC\,2631 and 
       Cederblad 110/111 in the Chamaleon I dark cloud; (vii) Comer\'on et al. (1993
        and references therein) for the $\rho$ Ophiuchi IR cluster. The larger E(B-V)$_{\rm FIR}$ values occur for IR star clusters, while the T Tauri groups tend to be associated with lower reddening values. This must reflect the need of higher dust densities for the formation of star clusters and massive stars, conditions which occur in the cores of Giant Molecular Clouds in contrast to less massive dark clouds (e.g. Comer\'on et al. 1993, Carpenter et al. 1997).

For the infrared photometric reddening comparisons with E(B-V)$_{\rm FIR}$ we adopt
a total to selective extinction ratio $R = \frac {A_{\rm V}}{E(B-V)} = 3.1$. When the original studies do not express their results in terms of $A_V$, we adopt the ratios $\frac {A_{\rm J}}{A_{\rm V}} = 0.276$, $\frac {A_{\rm H}}{A_{\rm V}} = 0.176$ and $\frac {A_{\rm K}}{A_{\rm V}} = 0.112$ from Schlegel et al. (1998).

JHK photometry of embedded stars in NGC\,2024 (Comer\'on et al. 1996) indicates an average E(B-V) $\approx 14.5$,  lower than that given by E(B-V)$_{\rm FIR}$ (Table 1) which might be accounted for if the sources are not the deepest embedded ones and if there still is a considerable amount of dust in the back half of the cloud. Another possibility is that hot stars heat the cloud core beyond the upper limit (21$^{\circ}$K) available in Schlegel et al.'s temperature maps.

HK photometry of deeply embedded stars in OMC-2 (Johnson et al. 1990) provided an average $E(B-V) = 7.7$, so that the reddening through the whole cloud should be twice as much. These values are consistent with  E(B-V)$_{\rm FIR} = 11.4$ (Table 1).

JHKL photometry  of a deeply embedded source in OMC-1 (Minchin et al. 1991) indicated  A$_{\rm L} = 6$ and A$_{\rm K} > 9$ which implied A$_{\rm V} \approx 200$ or 90 depending on models with large and small dust grain respectively. The latter values convert to E(B-V) = 64.5 or 29.0 respectively, which bracket that from E(B-V)$_{\rm FIR} = 50.9$ (Table 1), in this case favouring the large grain model.

JHK measurements of stars in the Mon R2 IR cluster (Carpenter et al. 1997) indicated that most of the embedded stars are in the range $ 3.3 \le E(B-V) \le 4.5$ somewhat higher than E(B-V)$_{\rm FIR} = 2.5$.

McGregor et al. (1994) obtained JHK photometry of a star projected close to the center of the reflection nebula Ced110 in Chamaleon I. They derived E(B-V) $\approx 2.4$ or $5.6$ depending on the assumed spectral type. In this direction E(B-V)$_{\rm FIR}$ = 3.25, in reasonable agreement. 

Comer\'on et al. (1993) estimated from JHK photometry that deeply embedded sources in the $\rho$ Oph dark cloud core have $E(B-V) > 16$ and that field stars in the background of the cloud are possibly affected by E(B-V) $\approx 22.5$. These values bracket that of E(B-V)$_{\rm FIR}$ (Table 1) which is  sensitive to reddening arising from the cloud as a whole.

Finally, we compare reddening values for the old/young components of the galactic nucleus (Catchpole et al. 1990, Krabbe et al. 1991), and for the two young star clusters projected close to the nucleus which contain WR stars (Quintuplet cluster = AFGL2004 and the Arches cluster = Object 17, respectively Glass et al. 1990, Nagata et al. 1995). The galactic center reddening is E(B-V) $\approx 9.7$, that for the Quintuplet cluster E(B-V) $\approx 7.1$ and that for the Arches cluster E(B-V)$\approx 10.6$. The E(B-V)$_{\rm FIR}$  values in these directions are exceedingly high  (E(B-V)$_{\rm FIR} \approx 98-100$). High values are expected since in this direction there is an integrated dust contribution behind the nucleus out to the disk edge. In the foreground of the nucleus we have contributions from the central molecular zone, the molecular ring and the four arms between the sun and the galactic nucleus. By symmetry arguments considering that there are three extra arms in the background outside the solar circle Schlegel et al.'s reddening values should not exceed E(B-V) $_{\rm FIR} \approx 25$. A possible explanation is that the dust temperature in the nucleus and surroundings is significantly higher than those in Schlegel et al.'s temperature map. Indeed the possibility of non-thermal photon flux and existence of three massive young clusters may account for the dust heating.

We conclude that Schlegel et al.'s reddening values are in general consistent with values obtained from infrared photometry of embedded sources in dark clouds. Some discrepancies for infrared star clusters in the cores of molecular clouds might be explained by dust heated a few degrees above  21$^{\circ}$K. As pointed out by Schlegel et al. (1998) the same dust column density provides a factor of 5 in 100 $\mu$m flux when heated from 17$^{\circ}$K to 21$^{\circ}$K which are the temperature extremes considered in their study. For the Central Molecular Zone in the Galaxy E(B-V) $_{\rm FIR}$  values appear to be exceedingly high so that the dust temperature near the galactic nucleus must be higher. Indeed, assuming that E(B-V)$_{\rm FIR}  \approx 25$ the dust temperature implied is $T \approx 26$$^{\circ}$K.

\subsection{Dust layer height from nearby dark clouds}  
    
    In the following we calculate the distance from the galactic plane (Z) 
    of the nearby dust complexes to estimate the dust layer height to which
    significant reddening is expected, at least in the solar neighbourhood.
    
   Considering that the sun appears to be located 15 pc above the galactic 
   plane (Cohen 1995, 
Hammersley et al. 1995), the corrected distance from the galactic plane $Z'$ is

\begin{equation}
Z' = d_{\rm sun} \sin b + 15,
\end{equation}
where d$_{\rm sun}$ is the object distance from the sun in pc.
   
   Hipparcos distances of the Orion and $\rho$ Ophiuchi complexes 
   are respectively d$_{\rm sun}$ = 490 pc and d$_{\rm sun}$ = 125 pc (de Zeeuw et al. 1999), which imply distances
   from the plane $Z'$ = -148 pc and  $Z'$ = 59 pc respectively. The distance to Monoceros R2 is
   d$_{\rm sun}$ = 830 pc (Carpenter et al. 1997) and $Z'$ = -166 pc. Finally, for 
   Chamaleon I d$_{\rm sun}$ = 140 pc (Lawson et al. 1996) and $Z'$ = -21 pc.
    Of these well-studied high latitude clouds Mon R2 and the Orion complex
     are intrinsically distant enough from the plane to estimate the 
     dust layer height. The centers of the complexes imply $|Z'| \approx 150$ pc. 
     Since dust in the Orion complex appears to absorb significantly 
     (E(B-V)$_{\rm FIR} \approx$ 0.20) to at least {\it b} = -25$^{\circ}$ -- see in Fig. 2 the
     b profiles for OMC-1 and Mon R2 (which includes Orion complex parts at higher
     latitudes), we adopt a  dust layer height of 200 pc 
     for the subsequent discussions.

\begin{table}
\caption[]{Dust emission reddening in selected regions}
\begin{scriptsize}
\label{tab1}
\renewcommand{\tabcolsep}{0.9mm}
\begin{tabular}{ccccc}
\hline\hline
$\ell$&{\it b}&Object&E(B-V)& C\\
$(^{\circ})$&$(^{\circ})$&&FIR&\\
\hline
192.45& -11.31&    Barnard 30 T Tau Group&           1.7& 1\\
196.82& -10.55&    Barnard 35 T Tau Group&           0.3& 1\\
205.17& -14.13&    NGC\,2071 IR Cluster&              20.5& 2\\ 
205.33& -14.31&    M\,78(NGC\,2068) IR Cluster&                33.6& 2\\ 
206.47& -16.32&    NGC\,2024 IR Cluster&              44.9& 2\\ 
206.85& -16.53&    NGC\,2023 IR Cluster&               2.1& 2\\ 
208.55& -19.19&    OMC-3 IR Cluster&                 7.8& 3\\
208.82& -19.24&    OMC-2 IR Cluster&                11.4& 4\\
208.99& -19.38&    OMC-1 IR Cluster&                50.9& 5\\
212.43& -18.99&    L\,1641 South IR Cluster&          10.5& 6\\
210.08& -19.83&    HH\,34/KMS\,12 IR T group&            2.3& 6\\
210.09& -19.58&    L\,1641 North IR cluster&           2.2& 6\\
210.43& -19.73&    V\,380 Ori IR cluster&              4.8& 6\\
210.80& -19.50&    KMS\,35 IR cluster&                 3.1& 6\\
210.91& -19.33&    KMS\,36 IR T Tau Group&             4.1& 6\\
210.97& -19.33&    L\,1641C IR T Tau Group&            4.6& 6\\
212.23& -19.36&    CK T Tau Group&                   3.9& 6\\
213.69& -12.60&    Mon R2 IR Cluster&                2.5& 7\\
297.02& -14.92&    IC\,2631(Ced 112) T Tau Group&      6.2& 8\\
297.35& -15.75&    Ced\,110/111 T Tau Group&           2.0& 8\\ 
352.95&  16.96&    $\rho$ Oph Dk Cloud IR Cluster&  18.6&9\\
\hline
\end{tabular}
\end{scriptsize}
\begin{list}{}
\item  Notes to Table 1 - Column 5 referred a cloud complex: 1- Orion Head, 2- Orion:in Giant Molecular Cloud  L\,1630, 3- Orion Molecular Cloud 3, 4- Orion Molecular Cloud 2, 5- Orion Molecular Cloud 1, 6- Orion: in Giant Molecular Cloud L\,1641, 7- Monoceros R2, 8- Chamaleon I Dark Cloud, 9- $\rho$ Ophiuchi Cloud
\end{list}
\end{table}

\section{Reddening Values  in star cluster directions}

Since star clusters span a wide range of galactic latitudes and distances from the sun, both within and outside the dust layer, they are ideal targets for comparison of Schlegel et al.'s reddening values with those derived from the stellar content methods. In the present section we compile reddening values of  globular clusters and intermediate age open clusters. Owing to their higher ages they probe the interstellar medium  without being physically related to the dust complexes, except for the possibility of interactions.

\subsection{Globular clusters}

 Harris (1996) compiled parameters for the 147 Milky Way globular clusters, keeping an updated version in the Web interface

{\em http://physun.physics.mcmaster.ca/Globular.html}.

In previous compilations, e.g. Webbink (1985), many globular clusters had scanty or no  information. Colour-magnitude diagrams (CMD) based on CCD observations are now almost complete for these objects as a consequence of recent efforts, especially for the reddened low latitude globular clusters in crowded fields (e.g. Ortolani et al. 1995a, Barbuy et al. 1998a and references therein).  Just to illustrate the progress achieved we mention Terzan 3 for which Webbink (1985) provided $E(B-V) = 0.32$ based on the cosecant law, but the  CMD showed a considerably higher reddening $E(B-V) = 0.72$ (Barbuy et al. 1998b).

\begin{table}
\caption[]{Properties of Galactic globular clusters}
\begin{scriptsize}
\label{tab2}
\renewcommand{\tabcolsep}{0.9mm}
\renewcommand{\arraystretch}{0.8}
\begin{tabular}{lcccccc}
\hline\hline
Name&$\ell$&{\it b}&d$_{\rm sun}$&E(B-V)&E(B-V)&$\beta$\\
&($^{\circ})$&($^{\circ})$&(kpc)&FIR&&E(B-V)\\
\hline
             47\,Tuc,NGC\,104 &                305.90 &                -44.89 &                   4.5 &                  0.03 &                  0.04 &                 -0.01 \\
                   NGC\,288 &                152.28 &                -89.38 &                   8.3 &                  0.01 &                  0.03 &                 -0.02 \\
                   NGC\,362 &                301.53 &                -46.25 &                   8.5 &                  0.03 &                  0.05 &                 -0.02 \\
                  NGC\,1261 &                270.54 &                -52.13 &                  16.4 &                  0.01 &                  0.01 &                  0.00 \\
                 Palomar\,1 &                130.07 &                 19.03 &                  10.9 &                  0.20 &                  0.15 &                  0.05 \\
                   AM\,1,E\,1 &                258.36 &                -48.47 &                 121.9 &                  0.01 &                  0.00 &                  0.01 \\
      Eridanus,ESO551-SC1 &                218.11 &                -41.33 &                  90.2 &                  0.02 &                  0.02 &                  0.00 \\
                 Palomar\,2 &                170.53 &                 -9.07 &                  27.6 &                  1.21 &                  1.24 &                 -0.03 \\
                  NGC\,1851 &                244.51 &                -35.04 &                  12.1 &                  0.04 &                  0.02 &                  0.02 \\
              M79,NGC\,1904 &                227.23 &                -29.35 &                  12.9 &                  0.04 &                  0.01 &                  0.03 \\
                  NGC\,2298 &                245.63 &                 -16.0 &                  10.7 &                  0.22 &                  0.14 &                  0.08 \\
                  NGC\,2419 &                180.37 &                 25.24 &                  84.2 &                  0.06 &                  0.11 &                 -0.05 \\
        Pyxis,Weinberger\,3 &                261.32 &                  7.00 &                  39.4 &                  0.32 &                  0.21 &                  0.11 \\
                  NGC\,2808 &                282.19 &                -11.25 &                   9.3 &                  0.23 &                  0.23 &                  0.00 \\
             E\,3,ESO37-SC1 &                292.27 &                -19.02 &                   4.3 &                  0.34 &                  0.30 &                  0.04 \\
                 Palomar\,3 &                240.14 &                 41.86 &                  92.7 &                  0.04 &                  0.04 &                  0.00 \\
                  NGC\,3201 &                277.23 &                  8.64 &                   5.2 &                  0.26 &                  0.21 &                  0.05 \\
                 Palomar\,4 &                202.31 &                 71.80 &                 109.2 &                  0.02 &                  0.01 &                  0.01 \\
                  NGC\,4147 &                252.85 &                 77.19 &                  19.3 &                  0.03 &                  0.02 &                  0.01 \\
                  NGC\,4372 &                300.99 &                 -9.88 &                   5.8 &                  0.56 &                  0.39 &                  0.17 \\
                   Rup\,106 &                300.89 &                 11.67 &                  21.2 &                  0.17 &                  0.20 &                 -0.03 \\
              M\,68,NGC\,4590 &                299.63 &                 36.05 &                  10.2 &                  0.06 &                  0.05 &                  0.01 \\
                  NGC\,4833 &                303.61 &                 -8.01 &                   6.0 &                  0.33 &                  0.33 &                  0.00 \\
              M\,53,NGC\,5024 &                332.96 &                 79.76 &                  18.3 &                  0.03 &                  0.02 &                  0.01 \\
                  NGC\,5053 &                335.69 &                 78.94 &                  16.4 &                  0.02 &                  0.04 &                 -0.02 \\
      $\omega$Cen,NGC\,5139 &                309.10 &                 14.97 &                   5.3 &                  0.14 &                  0.12 &                  0.02 \\
               M\,3,NGC\,5272 &                 42.21 &                 78.71 &                  10.4 &                  0.01 &                  0.01 &                  0.00 \\
                  NGC\,5286 &                311.61 &                 10.57 &                  11.0 &                  0.29 &                  0.24 &                  0.05 \\
                      AM\,4 &                320.28 &                 33.51 &                  29.9 &                  0.05 &                  0.04 &                  0.01 \\
                  NGC\,5466 &                 42.15 &                 73.59 &                  17.0 &                  0.02 &                  0.00 &                  0.02 \\
                  NGC\,5634 &                342.21 &                 49.26 &                  25.9 &                  0.06 &                  0.05 &                  0.01 \\
                  NGC\,5694 &                331.06 &                 30.36 &                  34.7 &                  0.10 &                  0.09 &                  0.01 \\
                   IC\,4499 &                307.35 &                -20.47 &                  18.9 &                  0.22 &                  0.23 &                 -0.01 \\
                  NGC\,5824 &                332.55 &                 22.07 &                  32.0 &                  0.17 &                  0.13 &                  0.04 \\
                 Palomar\,5 &                  0.85 &                 45.86 &                  23.2 &                  0.06 &                  0.03 &                  0.03 \\
                  NGC\,5897 &                342.95 &                 30.29 &                  12.8 &                  0.14 &                  0.09 &                  0.05 \\
               M\,5,NGC\,5904 &                  3.86 &                 46.80 &                   7.5 &                  0.04 &                  0.03 &                  0.01 \\
                  NGC\,5927 &                326.60 &                  4.86 &                   7.6 &                  0.51 &                  0.45 &                  0.06 \\
           NGC\,5946,IC\,4550 &                327.58 &                  4.19 &                  12.8 &                  0.71 &                  0.54 &                  0.17 \\
         BH\,176,ESO224-SC8 &                328.41 &                  4.34 &                  13.4 &                  0.59 &                  0.69 &                 -0.10 \\
                  NGC\,5986 &                337.02 &                 13.27 &                  10.5 &                  0.34 &                  0.27 &                  0.07 \\
             Lyng\aa\,7,BH\,184 &                328.77 &                 -2.79 &                   6.7 &                  1.06 &                  0.72 &                  0.34 \\
           Palomar\,14,AvdB &                 28.75 &                 42.18 &                  73.9 &                  0.03 &                  0.04 &                 -0.01 \\
              M\,80,NGC\,6093 &                352.67 &                 19.46 &                  10.0 &                  0.21 &                  0.18 &                  0.03 \\
               M\,4,NGC\,6121 &                350.97 &                 15.97 &                   2.2 &                  0.50 &                  0.36 &                  0.14 \\
                  NGC\,6101 &                317.75 &                -15.82 &                  15.3 &                  0.10 &                  0.05 &                  0.05 \\
                  NGC\,6144 &                351.93 &                 15.70 &                  10.3 &                  0.71 &                  0.32 &                  0.39 \\
                  NGC\,6139 &                342.37 &                  6.94 &                  10.1 &                  0.90 &                  0.75 &                  0.15 \\
                  Terzan\,3 &                345.08 &                  9.19 &                   6.5 &                  0.76 &                  0.72 &                  0.04 \\
             M\,107,NGC\,6171 &                  3.37 &                 23.01 &                   6.4 &                  0.45 &                  0.33 &                  0.12 \\
    ESO452-SC11,C1636-283 &                351.91 &                 12.10 &                   7.8 &                  0.52 &                  0.49 &                  0.03 \\
              M\,13,NGC\,6205 &                 59.01 &                 40.91 &                   7.7 &                  0.02 &                  0.02 &                  0.00 \\
                  NGC\,6229 &                 73.64 &                 40.31 &                  30.7 &                  0.02 &                  0.01 &                  0.01 \\
              M\,12,NGC\,6218 &                 15.72 &                 26.31 &                   4.9 &                  0.17 &                  0.19 &                 -0.02 \\
                  NGC\,6235 &                358.92 &                 13.52 &                  10.0 &                  0.42 &                  0.36 &                  0.06 \\
              M\,10,NGC\,6254 &                 15.14 &                 23.08 &                   4.4 &                  0.29 &                  0.28 &                  0.01 \\
            NGC\,6256,BH\,208 &                347.79 &                  3.31 &                   6.4 &                  1.72 &                  1.03 &                  0.69 \\
                Palomar\,15 &                 18.87 &                 24.30 &                  44.6 &                  0.40 &                  0.40 &                  0.00 \\
              M\,62,NGC\,6266 &                353.58 &                  7.32 &                   6.9 &                  0.46 &                  0.47 &                 -0.01 \\
              M\,19,NGC\,6273 &                356.87 &                  9.38 &                   8.7 &                  0.31 &                  0.37 &                 -0.06 \\
                  NGC\,6284 &                358.35 &                  9.94 &                  14.7 &                  0.31 &                  0.28 &                  0.03 \\
                  NGC\,6287 &                  0.13 &                 11.02 &                   8.5 &                  0.81 &                  0.60 &                  0.21 \\
                  NGC\,6293 &                357.62 &                  7.83 &                   8.8 &                  0.62 &                  0.41 &                  0.22 \\
                  NGC\,6304 &                355.83 &                  5.38 &                   6.1 &                  0.52 &                  0.52 &                  0.00 \\
                  NGC\,6316 &                357.18 &                  5.76 &                  11.0 &                  0.73 &                  0.51 &                  0.22 \\
              M\,92,NGC\,6341 &                 68.34 &                 34.86 &                   8.2 &                  0.02 &                  0.02 &                  0.00 \\
                  NGC\,6325 &                  0.97 &                  8.00 &                   9.6 &                  0.95 &                  0.89 &                  0.06 \\
               M\,9,NGC\,6333 &                  5.54 &                 10.70 &                   8.2 &                  0.43 &                  0.38 &                  0.05 \\
                  NGC\,6342 &                  4.90 &                  9.73 &                   8.6 &                  0.52 &                  0.46 &                  0.06 \\
                  NGC\,6356 &                  6.72 &                 10.22 &                  15.2 &                  0.31 &                  0.28 &                  0.03 \\
                  NGC\,6355 &                359.58 &                  5.43 &                   7.2 &                  1.21 &                  0.75 &                  0.46 \\
                  NGC\,6352 &                341.42 &                 -7.17 &                   5.7 &                  0.35 &                  0.21 &                  0.14 \\
                   IC\,1257 &                 16.53 &                 15.14 &                  25.0 &                  0.80 &                  0.73 &                  0.07 \\
              Terzan\,2,HP\,3 &                356.32 &                  2.30 &                   6.6 &                  2.38 &                  1.46 &                  0.92 \\
                  NGC\,6366 &                 18.41 &                 16.04 &                   3.6 &                  0.75 &                  0.71 &                  0.04 \\
              Terzan\,4,HP\,4 &                356.02 &                  1.31 &                   7.3 &                  6.34 &                  2.26 &                  4.08 \\
                HP\,1,BH\,229 &                357.42 &                  2.12 &                   6.4 &                  2.24 &                  1.15 &                  1.09 \\
                  NGC\,6362 &                325.55 &                -17.57 &                   8.1 &                  0.07 &                  0.08 &                 -0.01 \\
\hline 
\end{tabular}
\end{scriptsize}
\begin{list}{Notes to Table 2:}
\item  $^1$in Sagittarius Dwarf (Ibata et al. 1994, Da Costa \& Armandroff 1999); $^2$Barbuy et al. (1998a); $^3$Bica et al. (1998);  $^4$Ortolani et al. (1999a); $^5$Barbuy et al. (1998b); $^6$Ortolani et al. (1999b); $^7$ Barbuy et al. (1999); $^8$ Ortolani et al. (1999c);$^9$ Kaisler et al. 1997; $^{10}$ Rosino et al. (1997); $^{11}$Ortolani et al. (1993); $^{12}$Ortolani et al. (1995b); $^{13}$Bica et al. (1995); $^{14}$ Ortolani et al. (1998). 
\end{list}
\end{table}

\begin{table}
%\caption[]{}
\begin{scriptsize}
\label{tab2}
\renewcommand{\tabcolsep}{0.9mm}
\renewcommand{\arraystretch}{0.8}
\begin{tabular}{lcccccc}
\hline\hline
Name&$\ell$&{\it b}&d$_{\rm sun}$&E(B-V)&E(B-V)&$\beta$\\
&($^{\circ}$)&($^{\circ}$)&(kpc)&FIR&&E(B-V)\\
\hline
 Liller\,1             &354.84&  -0.16&  7.4$^2$&11.57&2.95$^{2,3}$ 	 & 8.92 \\
 NGC\,6380, Ton\,1      &350.18&  -3.42&  9.8$^14$&1.54&1.11$^{3,14}$ 	 & 0.43 \\
 Terzan\,1, HP\,2       &357.57&   1.00&  5.2$^4$&7.05&2.38$^4$ 	 & 4.67 \\
 Tonantzintla\,2,Pi\c{s}\,26  &350.80&  -3.42&  6.4$^{13}$&1.65&1.23$^{3,13}$ 	 & 0.42 \\
 NGC\,6388             &345.56&  -6.74& 11.5&0.41&0.40 	 & 0.01 \\
 M\,14, NGC\,6402       & 21.32&  14.81&  8.9&0.48&0.60 	 &-0.12 \\
 NGC\,6401             & 3.45 &  3.98 &12.0$^7$&0.96&0.53$^7$  	 & 0.43 \\
 NGC\,6397             &338.17& -11.96&  2.3&0.19&0.18 	 & 0.01 \\
 Palomar\,6            &  2.09&   1.78&  6.4$^2$&1.76&1.31$^{2,3}$ 	 & 0.45 \\
 NGC\,6426             & 28.09&  16.23& 20.4&0.35&0.36 	 &-0.01 \\
 Djorgovski\,1         &356.67&  -2.48&  5.6$^2$&2.13&1.68$^{2,3}$ 	 & 0.45 \\
 Terzan\,5,Terzan\,11   &  3.81&   1.67&  3.6&4.27&2.50 	 & 1.77 \\
 NGC\,6440             &  7.73&  3.80&  8.4&1.14&1.07 	 & 0.07 \\
 NGC\,6441             &353.53& -5.01& 11.2&0.61&0.44 	 & 0.17 \\
 Terzan\,6,HP\,5        &358.57& -2.16&  5.4$^2$&2.67&2.13$^{2,3}$& 0.54 \\
 NGC\,6453             &355.72& -3.87& 11.2&0.67&0.61 	 & 0.06 \\
 UKS\,1                &  5.12&  0.76&  7.4$^2$&6.81&3.10$^{2,3}$ 	 & 3.71 \\
 NGC\,6496             &348.02&-10.01& 11.5&0.23&0.15 	 & 0.08 \\
 Terzan\,9             &  3.60& -1.99&4.9$^6$  &2.75&1.78$^{3,6}$   	 & 0.97 \\
 ESO456-SC38, Djorg\,2 &  2.76& -2.51&5.5$^2$  &1.19&0.72$^{2,3}$   	 & 0.47 \\
 NGC\,6517             & 19.23&  6.76& 10.8&1.21&1.08 	 & 0.13 \\
 Terzan\,10            &  4.42& -1.86&  4.8$^2$&5.13&2.41$^2$ 	 & 2.72 \\
 NGC\,6522             &  1.02& -3.93&  6.1$^2$&0.57&0.55$^2$ 	 & 0.02 \\
 NGC\,6535             & 27.18& 10.44&  6.7&0.41&0.34 	 & 0.07 \\
 NGC\,6528             &  1.14& -4.17&  7.8$^2$&0.75&0.52$^2$ 	 & 0.23 \\
 NGC\,6539             & 20.80&  6.78&  8.4&1.09&0.97 	 & 0.12 \\
 NGC\,6540, Djorg 3    &  3.29& -3.31&  3.0$^2$&0.64&0.60$^2$ 	 & 0.04 \\
 NGC\,6544             &  5.84& -2.20&  2.6&1.71&0.73 	 & 0.98 \\
 NGC\,6541             &349.29&-11.18&  7.0&0.16&0.14 	 & 0.02 \\
 NGC\,6553             & 5.25 &-3.02 & 5.1$^2$ &1.36&0.70$^2$  	 & 0.66 \\
 NGC\,6558             &  0.20& -6.03&  7.4&0.46&0.44 	 & 0.02 \\
 IC\,1276, Palomar\,7   & 21.83&  5.67&  4.0$^5$&1.40&1.16$^5$ 	 & 0.24 \\
 Terzan\,12            &  8.36& -2.10&  3.4$^{14}$&2.85&1.93$^{3,14}$ 	 & 0.92 \\
 NGC\,6569             &  0.48& -6.68&  8.7&0.42&0.56 	 &-0.14 \\
 NGC\,6584             &342.14&-16.41& 13.4&0.11&0.10 	 & 0.01 \\
 NGC\,6624             &  2.79& -7.91&  8.0&0.26&0.28 	 &-0.02 \\
 M\,28, NGC\,6626       &  7.80& -5.58&  5.7&0.48&0.43 	 & 0.05 \\
 NGC\,6638             &  7.90& -7.15&  8.4&0.41&0.40 	 & 0.01 \\
 M\,69, NGC\,6637       &  1.72&-10.27&  8.6&0.17&0.16 	 & 0.01 \\
 NGC\,6642             & 9.81& -6.44 & 7.7 &0.38&0.41  	 &-0.03 \\
 NGC\,6652             & 1.53&-11.38 & 9.6 &0.11&0.09  	 & 0.02 \\
 M\,22, NGC\,6656       &  9.89& -7.55&  3.2&0.33&0.34 	 &-0.01 \\
 Palomar\,8            & 14.10& -6.80& 12.9&0.41&0.32 	 & 0.09 \\
 M\,70, NGC\,6681       &  2.85&-12.51&  9.0&0.11&0.07 	 & 0.04 \\
 NGC\,6712             &25.35 &-4.32 & 6.9 &0.39&0.45  	 &-0.06 \\
 M\,54, NGC\,6715$^1$   &  5.61&-14.09& 27.2&0.15&0.14 	 & 0.01 \\
 NGC\,6717, Palomar\,9  & 12.88&-10.90&  7.1$^8$&0.25&0.22$^{3,8}$ 	 & 0.03 \\
 NGC\,6723             & 0.07&-17.30 & 8.8 &0.16&0.05  	 & 0.11 \\
 NGC\,6749, Be\,42      & 36.20& -2.20&  7.3$^{10,9}$&1.77&1.45$^{10,9}$ 	 & 0.32 \\
 NGC\,6752             &336.50&-25.63&  4.0&0.06&0.04 	 & 0.02 \\
 NGC\,6760             &36.11 &-3.92 & 7.4 &0.65&0.77  	 &-0.12 \\
 M\,56, NGC\,6779       & 62.66&  8.34& 10.1&0.25&0.20 	 & 0.05 \\
 Terzan\,7$^1$         &  3.39&-20.07& 23.2&0.09&0.07 	 & 0.02 \\
 Palomar\,10           & 52.44 & 2.72&  5.9$^9$&2.05&1.58$^{3,9}$ 	 & 0.47 \\
 Arp\,2$^1$            &  8.55&-20.78& 28.6&0.11&0.10 	 & 0.01 \\
 M\,55, NGC\,6809       &  8.80&-23.27&  5.4&0.14&0.07 	 & 0.07 \\
 Terzan\,8$^1$         &  5.76&-24.56& 26.0&0.15&0.12 	 & 0.03 \\
 Palomar\,11           & 31.81&-15.58& 12.9&0.23&0.34 	 &-0.11 \\
 M\,71, NGC\,6838       &	56.74& -4.56&  3.9&0.33&0.25 	 & 0.08 \\
 M\,75, NGC\,6864       & 20.30&-25.75& 18.8&0.15&0.16 	 &-0.01 \\
 NGC\,6934             & 52.10&-18.89& 17.4&0.11&0.09 	 & 0.02 \\
 M\,72, NGC\,6981       & 35.16&-32.68& 17.0&0.06&0.05 	 & 0.01 \\
 NGC\,7006             &63.77&-19.41 &41.5 &0.08&0.05  	 & 0.03 \\
 M\,15, NGC\,7078       &	65.01&-27.31& 10.3&0.11&0.10 	 & 0.01 \\
 M\,2, NGC\,7089        & 53.38&-35.78& 11.5&0.04&0.06 	 &-0.02 \\
 M\,30, NGC\,7099       & 27.18&-46.83&  8.0&0.05&0.03 	 & 0.02 \\
 Palomar\,12           & 30.51&-47.68& 19.1&0.04&0.02 	 & 0.02 \\
 Palomar\,13           & 87.10&-42.70& 26.9&0.11&0.05 	 & 0.06 \\
 NGC\,7492             & 53.39&-63.48& 25.8&0.04&0.00 	 & 0.04 \\
\hline
\end{tabular}
\end{scriptsize}
\begin{list}{}
\item Second half of Table 2
\end{list}
\end{table}

Table 2 lists the galactic globular cluster, as follows: (1) name of object, (2) and (3) galactic coordinates, (4) distance from the sun, (5) reddening derived from dust 100 $\mu$m emission E(B-V)$_{\rm FIR}$, (6) E(B-V) derived from the light emitted by the cluster members,  and (7) $\beta$E(B-V) which is the difference between E(B-V)$_{\rm FIR}$ and  E(B-V) (Sect.4.1). The E(B-V)$_{\rm FIR}$ values were obtained from  Schlegel et al.'s reddening maps using the cluster galactic coordinates. Reddening and distance values are from Harris' (1996) compilation as updated to June 22 1999, except for low latitude globular clusters which come from the CMD studies indicated in the Table notes. 
Low latitude globular clusters have also been studied in detail via integrated spectral distribution in the near IR ($7000 <\lambda < 10000$\,\AA), which also is a direct estimator of the reddening affecting the stellar content (Bica et al. 1998). For these clusters, reddening values derived spectroscopically were also considered in Table 2 (see Table notes).

\subsection{Old open clusters}

 The old open clusters (700 Myr  or older), also usually referred to as Intermediate Age Clusters (IACs), are particularly suitable for  studying the galactic reddening at low and moderately high galactic latitude directions because the old disk is relatively thick (Friel 1995). They are numerous  for 90$^{\circ}<$ $\ell$ $<$ 270$^{\circ}$, thus complementary to the globular cluster sample which  in turn probes numerous lines of sight  towards the bulge.  We looked for old open clusters in  compilations (Janes \& Phelps 1994,
Friel 1995, Carraro et al. 1998), and individual clusters in the Open Cluster Database (Mermilliod 1996) as updated to November 1999 in the Web interface

{\em http://obswww.unige.ch/webda}.

We checked in the original references the CMD quality and the derived cluster parameters. In recent years the number of CCD photometric studies has been increasing steadily. They include clusters with CMD for the first time, CCD data on clusters previously  observed photographically, and finally clusters newly observed in the infrared (J and K bands). Just to mention some recent studies: NGC\,2204, NGC\,2477, Berkeley\,39 and Melotte\,66 (Kassis et al. 1997), Trumpler\,5 (Kaluzny 1998), Pi\c{s}mi\c{s}\,18, Pi\c{s}mi\c{s}\,19, NGC\,6005 and NGC\,6253 (Piatti et al. 1998a), Berkeley\,17 and Berkeley\,18 (Carraro et al. 1999), and ESO93-SC08 (Bica et al. 1999).

Janes \& Phelps' (1994) compilation included 72 IACs while the present sample has 103 entries. The Hyades were not included due to the proximity and large angular size. Table 3 lists the galactic old open clusters, as follows: (1) name of object, (2) and (3) galactic coordinates, (4) distance from the sun, (5) age, (6) E(B-V)$_{\rm FIR}$, (7) E(B-V),  and (8) $\beta$E(B-V). 
 
Figure 3 shows the sample properties. The histogram giving the distribution of old open clusters as a function of the distance from the sun. It  shows that the sample is probing the interstellar medium quite far, mostly in the range 1-5 kpc and in some cases as far as 10-14 kpc. The age histogram shows a steady decrease for older ages probably related to the dissolution rate of IACs. There occurs a peak at t $\approx$ 5 Gyr which was also present in previous compilations, and the present increased sample supports its significance. A possible interpretation for this peak would be a burst of star formation in the old disk.

\begin{table}
\caption[]{Properties of galactic old open clusters}
\begin{scriptsize}
\label{tab3}
\renewcommand{\tabcolsep}{0.9mm}
\renewcommand{\arraystretch}{0.8}
\begin{tabular}{lccccccc}
\hline\hline
Name&$\ell$&{\it b}&d$_{\rm sun}$&t&E(B-V)&E(B-V)&$\beta$\\
&($^{\circ}$)&($^{\circ}$)&(kpc)&(Myr)&FIR&&E(B-V)\\
\hline
Be\,81               &      34.66&  -1.95&	 3.00&	1000&	 2.92&  1.00&  1.92\\
IC\,4756,Mel-210     &  	 36.37&   5.25&	 0.44&	 750&	 0.76&  0.20&  0.56\\
NGC\,6802,Cr\,400      &  	 55.32 &  0.94&	 0.93&	1750&	 4.32&  0.83&  3.49\\
NGC\,6940,Mel-232    &  	 69.90&  -7.16&	 0.93&	 750&	 0.48&  0.24&  0.24\\
NGC\,6791,Be\,46       &  	 70.01&  10.95&	 4.24&	9000&	 0.15&  0.15&  0.00\\
NGC\,6819,Mel-223    &  	 73.97&   8.47&	 1.86&	2550&	 0.20&  0.24& -0.04\\
IC1311,Tr36        &  	 77.69&   4.27&	 4.38&	1200&	 0.65&  0.45&  0.20\\
NGC\,6811,Mel-222    &  	 79.44&  11.95&	 1.29&	 650&	 0.14&  0.13&  0.01\\
NGC\,6866,Mel-229    &  	 79.54&   6.86&	 1.49&	 650&	 0.77&  0.12&  0.65\\
Be\,54               &  	 83.12&  -4.14&	 3.80&	5600&	 1.19&  0.84&  0.35\\
NGC\,7044,Cr\,433      &  	 85.86&  -4.13&	 3.35&	1770&	 1.03&  0.67&  0.36\\
IC\,1369,Cr\,432       &  	 89.56&  -0.40&	 0.99&	1450&	 2.69&  0.60&  2.09\\
NGC\,6939,Mel-231    &  	 95.86&  12.31&	 1.20&	1800&	 0.39&  0.50& -0.11\\
NGC\,7226,Cr\,446      &  	101.42&  -0.59&	 2.56&	 600&	 1.03&  0.47&  0.56\\
Ki9                &  	101.43&  -1.82&	 4.56&	4400&	 0.61&  0.72& -0.11\\
NGC\,7142,Cr\,442      &  	105.42&   9.45&	 1.95&	5100&	 0.50&  0.40&  0.10\\
Ki\,19,Basel\,2       &  	110.57&   0.14&	 1.39&	1850&	 2.25&  0.48& 1.77\\
NGC\,7789,Mel-245    &  	115.48&  -5.35&	 1.84&	1520&	 0.41&  0.25&  0.16\\
Be\,99               &  	115.95&  10.11&	 4.90&	3150&	 0.48&  0.30&  0.18\\
Ki\,11               &  	117.15&   6.47&	 2.13&	5800&	 0.97&  1.00& -0.03\\
NGC\,7762,Mel-244    &  	117.18&   5.84&	 0.77&	1750&	 1.19&  0.81&  0.38\\
Be\,2                &  	119.70&  -2.31&	 5.25&	 800&	 0.93&  0.80&  0.13\\
NGC\,188,Mel-2       &  	122.77&  22.47&	 1.60&	4900&	 0.09&  0.09&  0.00\\
Ki\,2                &  	122.88&  -4.68&	 6.00&	5800&	 0.44&  0.31&  0.13\\
Cr463              &  	127.36&   9.56&	 0.35&	 700&	 0.73&  0.24&  0.49\\
IC\,166,Tom\,3         &  	130.07&  -0.19&	 3.08&	1200&	 1.15&  0.80&  0.35\\
Be\,64               &  	131.91&   4.59&	 3.88&	1000&	 1.07&  1.04&  0.03\\
NGC\,752,Mel-12      &  	137.17& -23.35&	 0.37&	1700&	 0.05&  0.04&  0.01\\
Be\,66               &  	139.41&   0.20&	 5.00&	4700&	 1.43&  1.25&  0.18\\
Ki\,5                &  	143.74&  -4.26&	 2.19&	 850&	 0.95&  0.78&  0.17\\
NGC\,1245,Mel-18     &  	146.63&  -8.92&	 2.80&	1080&	 0.31&  0.29&  0.02\\
NGC\,1193,Cr\,35       &  	146.80& -12.17&	 4.01&	4950&	 0.24&  0.12&  0.12\\
Ki\,7                &  	149.77&  -1.02&	 2.20&	 690&	 1.81&  1.25&  0.56\\
NGC\,1496,Cr\,44       &  	149.86&   0.14&	 1.23&	 630&	 1.42&  0.45&  0.97\\
NGC\,1798,Be16       &  	160.77&   4.83&	 3.82&	1450&	 0.60&  0.51&  0.09\\
Ki\,22,Be\,18          &  	163.62&   5.01&	 4.67&	4600&	 0.61&  0.47&  0.14\\
NGC\,2192,Mel-42     &  	173.41&  10.63&	 3.47&	1100&	 0.18&  0.20& -0.02\\
Be\,69               &  	174.43&  -1.79&	 2.86&	 890&	 0.84&  0.65&  0.19\\
Be\,17               &  	175.63&  -3.66&	 2.53&	9000&	 0.76&  0.64&  0.12\\
Ki\,8                &  	176.39&   3.11&	 3.35&	 800&	 0.84&  0.68&  0.16\\
Be\,19               &  	176.90&  -3.59&	 4.83&	3000&	 0.53&  0.40&  0.13\\
NGC\,1817,Cr\,60       &  	186.13& -13.12&	 1.97&	 950&	 0.33&  0.29&  0.04\\
NGC\,2158,Mel-40     &  	186.64&   1.78&	 4.09&	2200&	 0.75&  0.45&  0.30\\
Be\,21	           &     186.84&  -2.51&	 5.40&	2700&	 0.84&  0.69&  0.15\\
NGC\,2266,Mel-50     &     187.78&  10.27&	 3.38& 	 800&	 0.11&  0.10&  0.01\\
NGC\,2194,Mel-43     &  	197.26&  -2.33&	 2.65&	1000&	 0.82&  0.42&  0.40\\
Be\,29               &  	197.97&   8.02&	10.50&	4800&	 0.10&  0.21& -0.11\\
NGC\,2141,Cr\,79       &  	198.07&  -5.78&	 4.25&	2950&	 0.42&  0.33&  0.09\\
NGC\,2420,Mel-69     &  	198.11&  19.65&	 2.37&	2300&	 0.04&  0.03&  0.01\\
Be\,22               &  	199.80&  -8.04&	 5.05&	3300&	 0.65&  0.63&  0.02\\
Tr\,5,Cr\,105          &  	202.86&   1.06&	 2.60&	4900&	 0.95&  0.61&  0.34\\
NGC\,2355,Mel-63     &  	203.36&  11.80&	 2.20&	 800&	 0.14&  0.12&  0.02\\
Be\,20               &  	203.50& -17.28&	 8.27&	4600&	 0.21&  0.13&  0.08\\
NGC\,2236,Cr\,94       &  	204.37&  -1.68&	 3.21&	 930&	 0.96&  0.36&  0.60\\
NGC\,2395,Cr\,144      &  	204.62&  13.95&	 0.61&	1450&	 0.10&  0.07&  0.03\\
Praesepe,M\,44       & 	205.53&  32.52&	 0.17&	 800&	 0.03&  0.01&  0.02\\
NGC\,2112,Cr\,76        & 	205.90& -12.59&	 0.80&	5000&	 1.06&  0.50&  0.56\\
Biur\,7,Be\,31           &	206.25&   5.12&	 3.74&	3700&	 0.14&  0.24& -0.10\\
Biur\,8,Be\,32           	&207.93&   4.37&	 3.09&	4500&	 0.26&  0.16&  0.10\\
Cr\,110                &	209.64&  -1.97&	 1.95&	1400&	 1.12&  0.50&  0.62\\
Biur\,9,Be\,30           &	210.78&   2.86&	 2.34&	 900&	 0.63&  0.61&  0.02\\
NGC\,2324,Mel-59       &	213.39&   3.22&	 3.70&	 850&	 0.29&  0.13&  0.16\\
NGC\,2286,Cr\,117        &	215.31&  -2.29&	 2.19&	1600&	 1.26&  0.03&  1.23\\
M\,67,NGC\,2682          &	215.66&  31.94&	 0.82&	5300&	 0.03&  0.06& -0.03\\
Be\,39                 &	223.46&  10.08&	 4.40&	7400&	 0.12&  0.12&  0.00\\
NGC\,2204,Mel-44       &	226.01& -16.09&	 4.27&	2150&	 0.10&  0.09&  0.01\\
Haf\,8                 &	227.53&   1.35&	 1.18&	1430&	 0.66&  0.03&  0.63\\
Haf\,6                 &	227.85&   0.24&	 3.23&	 950&	 0.76&  0.43&  0.33\\
Mel-71,Cr\,155         &	228.95&   4.51&	 2.69&	 950&	 0.26&  0.06&  0.20\\
NGC\,2360,Mel-64       &	229.79&  -1.40&	 1.30&	1250&	 0.82&  0.08&  0.74\\
NGC\,2423,Mel-70       &	230.47&   3.54&	 0.70&	1200&	 0.32&  0.12&  0.20\\
NGC\,2506,Mel-80       &	230.60&   9.96&	 3.01&	2250&	 0.08&  0.09& -0.01\\
Tom\,1,Haf\,1            &	232.33&  -7.31&	 3.00&	1000&	 0.50&  0.40&  0.10\\
Tom\,2,Haf\,2            &	232.83&  -6.88&	 9.60&	2100&	 0.39&  0.25&  0.14\\
NGC\,2539,Mel-83       &	233.71&  11.10&	 1.48&	 600&	 0.07&  0.08& -0.01\\
NGC\,2243,Mel-46       &	239.48& -18.01&	 3.75&	6000&	 0.07&  0.05&  0.02\\
NGC\,2527,Cr\,174       & 	246.08&   1.85&	 0.61&	 850&	 0.47&  0.08&  0.39\\
NGC\,2533,Cr\,175       & 	247.80&   1.30&	 1.55&	1900&	 0.68&  0.01&  0.67\\
AM\,2,ESO368SC7       & 	248.12&  -5.87&	 8.35&	8300&	 0.73&  0.56&  0.17\\
NGC\,2627,Mel-87      & 	251.58&   6.65&	 1.91&	2800&	 0.15&  0.15&  0.00\\
NGC\,2477,Mel-78      & 	253.56&  -5.83&	 1.12&	 900&	 0.65&  0.29&  0.36\\
Pi\c{s}\,3                & 	257.86&   0.49&	 1.35&	2550&	 1.56&  1.35&  0.21\\
Pi\c{s}\,2                & 	258.85&  -3.33&	 2.84&	1700&	 2.27&  1.48&  0.79\\
Mel-66,Cr\,147        & 	259.56&  -14.24& 2.88&	5900&	 0.22&  0.17&  0.05\\
NGC\,2818A,Mel-96     & 	262.00&   8.60&	 3.93&	1050&	 0.19&  0.18&  0.01\\
NGC\,2660,Mel-92      & 	265.93&  -3.00&	 2.89&	1100&	 1.11&  0.37&  0.74\\
NGC\,1901,Bok\,1       & 	279.04& -33.64&	 0.42&	 832&	 0.33&  0.06&  0.27\\
NGC\,3680,Mel-106     & 	286.76&  16.93&	 0.90&	2200&	 0.09&  0.06&  0.03\\
ESO92SC18           & 	287.12&  -6.65&	 7.90&	5300&	 0.29&  0.26&  0.03\\
NGC\,3496,Cr\,237       & 	289.51&  -0.40&	 1.12&	 850&	 2.40&  0.50&  1.90\\
ESO93SC8            & 	293.50&  -4.04&   13.70&	4500&	 0.91&  0.64&  0.27\\
NGC\,3960,Mel-108     & 	294.36&   6.18&	 1.68&	 800&	 0.44&  0.29&  0.15\\
Harv\,6,Cr\,261         & 	301.68&  -5.53&	 2.38&	8500&	 0.45&  0.30&  0.15\\
AL\,1,ESO96SC4        & 	305.36&  -3.16&	 7.57&	 800&	 0.94&  0.72&  0.22\\
IC\,4291,Pi\c{s}\,18        & 	308.24&   0.32&	 2.24&	1200&	 5.79&  0.50&  5.29\\
Pi\c{s}\,19               & 	314.70&  -0.30&	 2.40&	1000&	12.52&  1.45& 11.07\\
NGC\,5823,Mel-131     & 	321.12&   2.45&	 0.71&	1650&	 2.04&  0.11&  1.93\\
NGC\,5822,Mel-130     & 	321.57&   3.59&	 0.83&	 850&	 0.98&  0.17&  0.81\\
NGC\,6005,Mel-138     & 	325.78&  -2.98&	 2.69&	1200&	 0.84&  0.45&  0.39\\
NGC\,6208,Cr\,313       & 	333.75 & -5.76&	 1.25&	2050&	 0.33&  0.15&  0.18\\
NGC\,6134,Mel-146     &    334.91&  -0.19&	 0.88&	1150&	25.66&  0.39& 25.27\\
NGC\,6253,Mel-156     &    335.45&  -6.25&	 1.51&	4000&	 0.35&  0.24&  0.11\\
IC\,4651,Mel-169       &   340.08 & -7.90& 0.82&	2400&	 0.25&  0.11&  0.14\\
\hline
\end{tabular}
\end{scriptsize}
\end{table}

\begin{figure} 
\resizebox{\hsize}{!}{\includegraphics{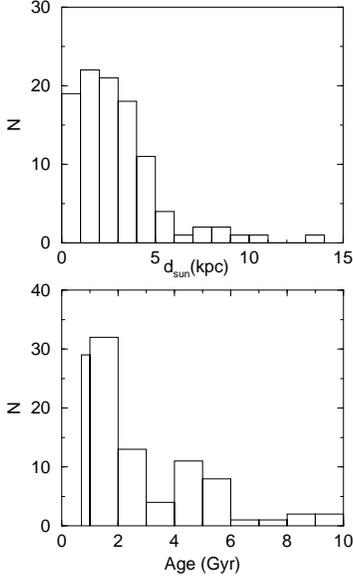}}
\caption[]{Distance (upper panel) and age (lower panel) histograms for old open clusters. In the age histogram the first bin considers clusters in the range 0.7 to 1 Gyr.}
\label{fig3}
\end{figure}

\section{Discussion}

\begin{figure} 
\resizebox{\hsize}{!}{\includegraphics{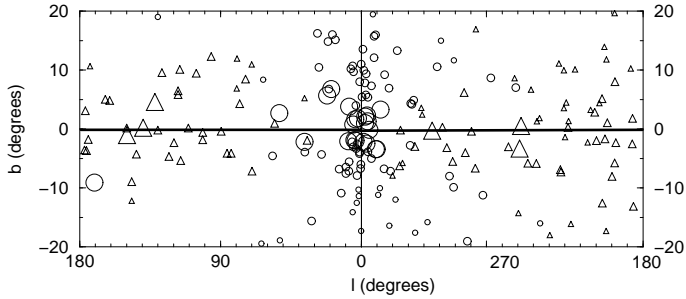}}
\caption[]{Angular distribution in galactic coordinates for globular clusters (circles) and old open clusters (triangles) within $|{\it b}| < 20^{\circ}$. The symbol size is proportional to reddening derived from the stellar content: (i) large corresponds to  E(B-V)$ > $ 1.0; (ii) intermediate  $ 0.2 \le E(B-V) \le 1.0$; and (iii) small E(B-V)$<$ 0.2.}
\label{fig4}
\end{figure}

The angular distribution  in galactic coordinates of globular  and old open  clusters for $|{\it b}| <$ 20$^{\circ}$  is shown in Fig.4, centred on the galactic nucleus direction. The two samples are complementary, globular clusters probe mostly the  galactic central regions while the old open clusters probe mostly the anticentre regions. The most frequent values for $|{\it b}| <$ 10$^{\circ}$ are intermediate ones ($ 0.2 \le E(B-V) \le 1.0$). For higher latitudes smaller values dominate. 

Reddening values for globular clusters in bulge regions often exceed  E(B-V) =  1. Terzan\,1, 4, 5, 6 and 10 exceed E(B-V) = 2, while Liller\,1 and  UKS\,1 have E(B-V)$\approx$ 3 (Table 2). The dust emission reddening  can be much larger, in some cases exceeding E(B-V)$_{\rm FIR}$ = 4 which occurs for  Terzan\,5, 10, 4, 1 and UKS\,1. The lowest galactic latitude globular cluster Liller\,1 has the highest value (E(B-V)$_{\rm FIR}$ = 11.57).  

For the old open clusters the largest reddening derived from the CMD occurs for Pi\c{s}mi\c{s}\,2 (E(B-V) = 1.48), and seven other clusters exceed E(B-V) = 1 (Table 3). Three clusters have dust emission reddening exceeding E(B-V)$_{\rm FIR}$ = 5 which are IC\,4291 (Pi\c{s}mi\c{s}\,18), Pi\c{s}mi\c{s}\,19 and finally NGC\,6134 with E(B-V)$_{\rm FIR}$ = 25.66. They are at extremely low latitudes and in directions not far from the galactic centre (Table 3) which can accumulate reddening from dust in several arms and the Molecular Ring (Sect.2).

\begin{figure*} 
\resizebox{\hsize}{!}{\includegraphics{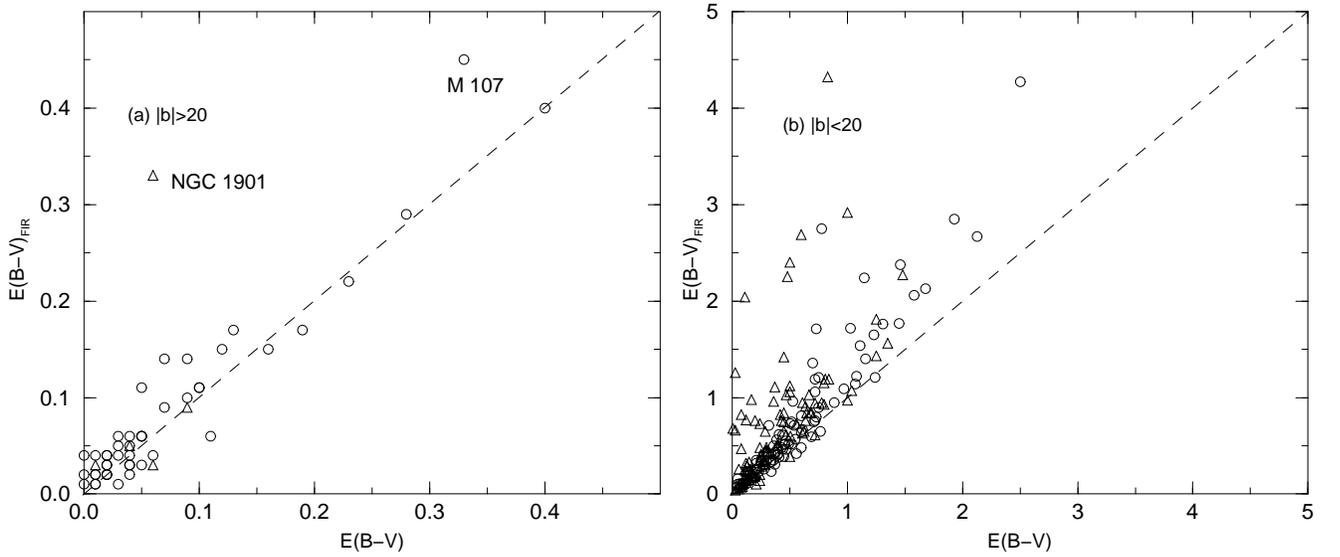}}
\caption[]{E(B-V)$_{\rm FIR}$ versus E(B-V) of the Galactic globular clusters (circles) and old open clusters (triangles). Panels: (a) high latitude objects  ($|b| > 20^{\circ}$), (b) low latitude ($|{\it b}| < 20^{\circ}$). The dashed line represents the identity function. Note that in panel (b) a few very reddened objects (E(B-V)$_{\rm FIR} > 5$) are beyond the figure limits (Tables 2 and 3).}
\label{fig5}
\end{figure*}

Figure 5 shows E(B-V)$_{\rm FIR}$ as a function of the reddening derived from the stellar content. Panel (a) contains the clusters with $|b| > 20^{\circ}$, presenting a good agreement between the reddening values, except the clusters M\,107 and NGC\,1901 which are discussed in Sect.4.1. The highest values at such latitudes are  E(B-V)$_{\rm FIR} \approx E(B-V) \approx 0.4$. Panel (b) contains the clusters with $|b| < 20^{\circ}$ where most points follow the identity function up to E(B-V)$\approx$ 1.0. However, an important fraction of points in the range $0< E(B-V) < 1$ has large deviations in the sense of higher E(B-V)$_{\rm FIR}$. For $E(B-V) > 1$ the points deviate systematically from the identity function in the sense that E(B-V)$_{\rm FIR}$ values are higher. A possible interpretation would be dust contributions for E(B-V)$_{\rm FIR}$ arising from the cluster background.

\subsection{$\beta$E(B-V): possibility of background reddening}

In order to check the possibility of background reddening in the directions of globular and old open clusters we define the difference 
 $\beta$E(B-V) = E(B-V)$_{\rm FIR}$ - E(B-V) (Tables 2 and 3, respectively). Figure 6  shows $\beta$E(B-V) histograms considering both samples together. For high latitude clusters ($|b| > 20^{\circ}$) we find a tight gaussian distribution suggesting an error distribution. We recall that E(B-V)$_{\rm FIR}$ uncertainties typically amount to 16\,\% (Schlegel et al. 1998). The gaussian peak is in the bin 0-0.02, indicating a small offset between the two reddening types with higher values for E(B-V)$_{\rm FIR}$. This can also be seen in Panel (a) of Fig.5 as a small systematic shift of the points with respect to the identity function. There are two deviating objects in Panel (a) of Fig.6 which are the globular cluster M\,107 (NGC\,6171) and the open cluster NGC\,1901. For M\,107 E(B-V)$_{\rm FIR}$ = 0.45 and E(B-V) = 0.33 (Table 3), thus $\beta$E(B-V) = 0.12. Recently Salaris \& Weiss (1997) derived E(B-V) = 0.38 from isochrone fitting on CCD photometry,  and they remarked that for this cluster values in the literature are in the range $0.30 < E(B-V) < 0.48$. We suspect that the positions of M\,107 in Figs. 5 and 6 reflect an uncertainty in the reddening derived from the stellar methods. On the other hand NGC\,1901 with E(B-V)$_{\rm FIR}$ = 0.33 and E(B-V) = 0.06 (Table 3) has the LMC disk as background (Sanduleak \& Philip 1968), so that the high $\beta$E(B-V) = 0.27 must reflect dust emission from LMC complexes.
 Panels (b) and (c) deal with $\beta$E(B-V) histograms for low latitude clusters ($|b| < 20^{\circ}$), respectively for $\beta E(B-V) < 1.5$ and high $\beta$E(B-V) values. Similarly to panel (a) there occurs in panel (b) a peak near zero (bin 0-0.04), which indicates that most points indeed closely follow the identity function (Panel (b) of Fig.5). Thus, Schlegel et al.'s (1998) reddening values at low latitudes agree with those derived from stellar data for about two thirds of the clusters. Finally, the histogram  for high $\beta$E(B-V) values (Panel (c)) shows 18 clusters with $\beta E(B-V) > 1.0$. These interesting objects together with the deviating clusters in Panel (b), typically with $\beta E(B-V) > 0.30$, are discussed in detail in Sect.4.4 for the possibility that their reddening values have an origin in the background dust.    

Since mass loss is            important in the last stages of red giant evolution dust accumulation in globular clusters is not unexpected. Cloudlets would contribute to differential reddening in CMDs as well as to 100 $\mu$m dust emission. Forte \& Mendez (1988) found evidence for dust within globular clusters. They studied ten southern globular clusters, in particular NGC\,362 and NGC\,6624, and detected by means of CCD imaging regions with light deficiency which was attributed to dark clouds with intrinsic extinctions close to A$_V$ = 2.5. Their sizes are on the order of tenths of a parsec and they occur near the  cluster nucleus.  We checked whether  Schlegel et al.'s reddening values are sensitive to internal dust contributions in the clusters NGC\,362 and NGC\,6624. We extracted E(B-V)$_{\rm FIR}$ values for a cross with 17 pixels in  Schlegel et al.'s maps (each pixel has 2.4$^{\prime}$ $\times$ 2.4$^{\prime}$). This cross samples the cluster main body and zones outside it, but still within the tidal radius (Trager et al. 1995). We noted fluctuations in E(B-V)$_{\rm FIR}$ not exceeding  0.01 and 0.02, respectively. We conclude that Schlegel et al.'s reddening values are not particularly sensitive to the cloudlets owing to the large pixel size and the  cloudlets' small covering factor. 

\begin{figure} 
\resizebox{\hsize}{!}{\includegraphics{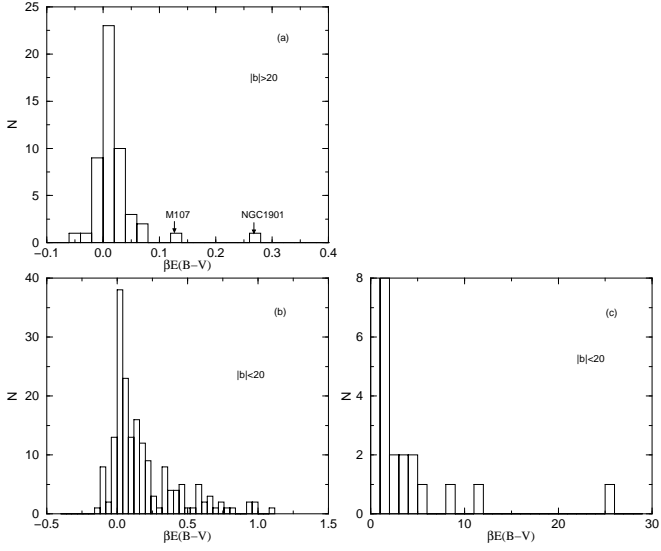}}
\caption[]{Histograms of $\beta$E(B-V) for the total sample of globular clusters and old open clusters: (a) high latitude clusters, (b) low latitude clusters for $\beta E(B-V) < 1.5$ and finally (c) low latitude clusters for the entire $\beta$E(B-V) range. The first bin in panel (c) is cropped for clarity purposes.}
\label{fig6}
\end{figure}

Star clusters within the dust layer as defined in Sect.2.2 ($|Z'| \approx 200$ pc) are expected to have significant differences between  E(B-V)$_{\rm FIR}$ and E(B-V) values. In order to study the behaviour of $\beta$E(B-V) we considered the cluster perpendicular distance to the Galactic 
plane ($Z'$) calculated with Eq.(1),  using the data in Tables 2 and 3 for the globular and old open clusters. 
Figure 7 shows  $\beta$E(B-V) as a function of $Z'$ for four $Z'$ ranges. Panel (a) shows objects up to 2 kpc from the Plane. Clearly, there is a large scatter of $\beta$E(B-V) values within the 200 pc dust layer, together with a significant wing which extends to $\approx 400$ pc. The scatter suggests that most of the  differences between reddening values derived from dust emission and the stellar content are due to dust clouds in the disk background of the clusters. Panels (b), (c) and (d) show the behaviour of $\beta$E(B-V) in the regions increasingly away from the disk.  In all three panels $\beta$E(B-V) values are small which is consistent with the fact that all these objects are halo globular clusters. The only strongly deviating object in Panel (b) is NGC6144 (Sect.4.4). In (b) and (c) the average $\beta$E(B-V) value is slightly positive corresponding to the small offset caused by higher E(B-V)$_{\rm FIR}$ values. Unless an extremely thin diffuse distribution occurs throughout the halo caused by e.g. cooling flows and/or debris from dwarf galaxies accreted by the Milky Way, this offset observed for halo globular clusters implies that either Schlegel et al.'s zero point is slightly overestimated or that intrinsic reference colors, spectral distributions and isochrones were exceedingly red. Finally, in Panel (d) the offset is not present, but the sample is small.

\begin{figure} 
\resizebox{\hsize}{!}{\includegraphics{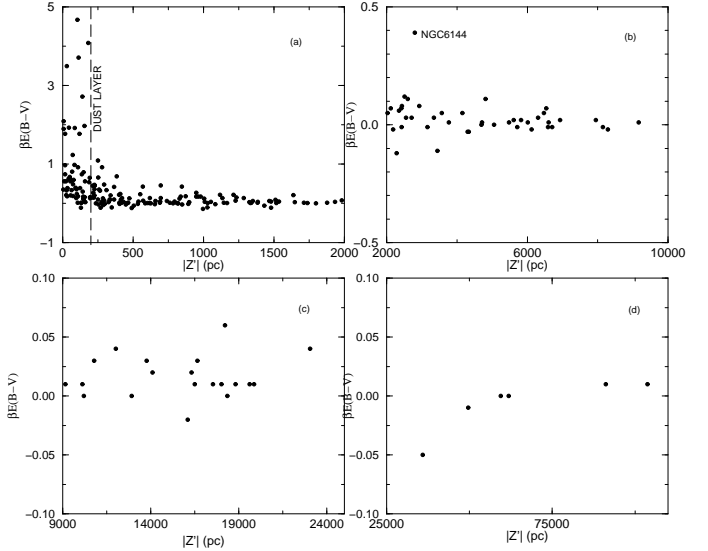}}
\caption[]{$\beta$E(B-V) for globular and old open  clusters as function of distance from the Plane $|Z'|$ in parsecs. The four panels show distinct ranges of distances from the galactic plane. Note that the $\beta$E(B-V) scale becomes amplified from (a) to (d). In panel (a) the dust layer of 200 pc is indicated and the values are cropped at $\beta$E(B-V) = 5.0. }
\label{fig7}
\end{figure}

Large differences $\beta$E(B-V) could be caused by the existence of dust clouds behind the clusters, primarily within the dust layer. In order to investigate this possibility we calculated cluster positions in the Galaxy and compared them to the assumed dust layer distribution.
The latitude  with respect to the true Galactic plane is given by

\begin{equation}
\tan {\it b'} = \frac{Z'}{d \sin {\it b}} \tan {\it b}. 
\end{equation}

Using this corrected latitude ${\it b'}$ it is possible to calculate the perpendicular 
distance of the cluster to the true Galactic plane $d'$, the distance along the cluster line of sight from the true Plane up to the dust layer edge  d$_{\rm layer}$, and for the clusters within the dust layer the path behind them d$_{\rm bck}$:
 
 \begin{equation}
d' = \frac{Z'}{\sin {\it b'}},\ 
d_{\rm layer} = \frac{200\,pc}{\sin {\it b'}},\ 
d_{\rm bck} = d_{\rm layer} - d'  
\end{equation}
where 200 pc refers to the conservative dust layer height, assumption of Sect.2.2. We also assumed a galactic disk radius  $R = 15$ kpc.

Figure 8 shows in panel (a) and in the blowup (b) E(B-V)$_{\rm FIR}$ as function of d$_{\rm layer}$. We note in panel (a) that objects with large E(B-V)$_{\rm FIR}$ values (e.g. NGC\,6134, Pi\c{s}mi\c{s}\,19, Liller\,1, IC\,4291) have a large path length within the layer.  This is expected since the reddening derived from dust emission (E(B-V)$_{\rm FIR}$) should integrate dust contributions along the whole path throughout to the disk edge (d$_{\rm layer}$). Panel (b) suggests two correlations with different slopes, which might indicate differences in the cumulative effect of dust emission due to a discrete distribution of dust clouds. In panels (c) and its blowup (d) the behaviour of $\beta$E(B-V) is shown as a function of the path behind the cluster up to the disk edge (d$_{\rm bck}$). Panel (c) suggests that large $\beta$E(B-V) values  for clusters like  NGC\,6134, Pi\c{s}mi\c{s}\,19, Liller\,1, IC\,4291 could arise from a dust cloud distribution behind the clusters since their directions and positions in the Galaxy imply a large background path  up to the disk edge. Panel (d) shows considerable scatter which might be due to different origins: (i) inhomogeneous distribution of dust clouds, (ii) considerable uncertainties, (iii) the assumptions of a dusty disk are not satisfactory. Notice that for the assumed dust layer height of 200 pc the clusters outside the dust layer (d$_{\rm bck} = 0$) have an considerable range of values $0 < \beta E(B-V) < 1$. This suggests that the Milky Way dust lane could be thicker (Sect.4.4).

\begin{figure*} 
\resizebox{\hsize}{!}{\includegraphics{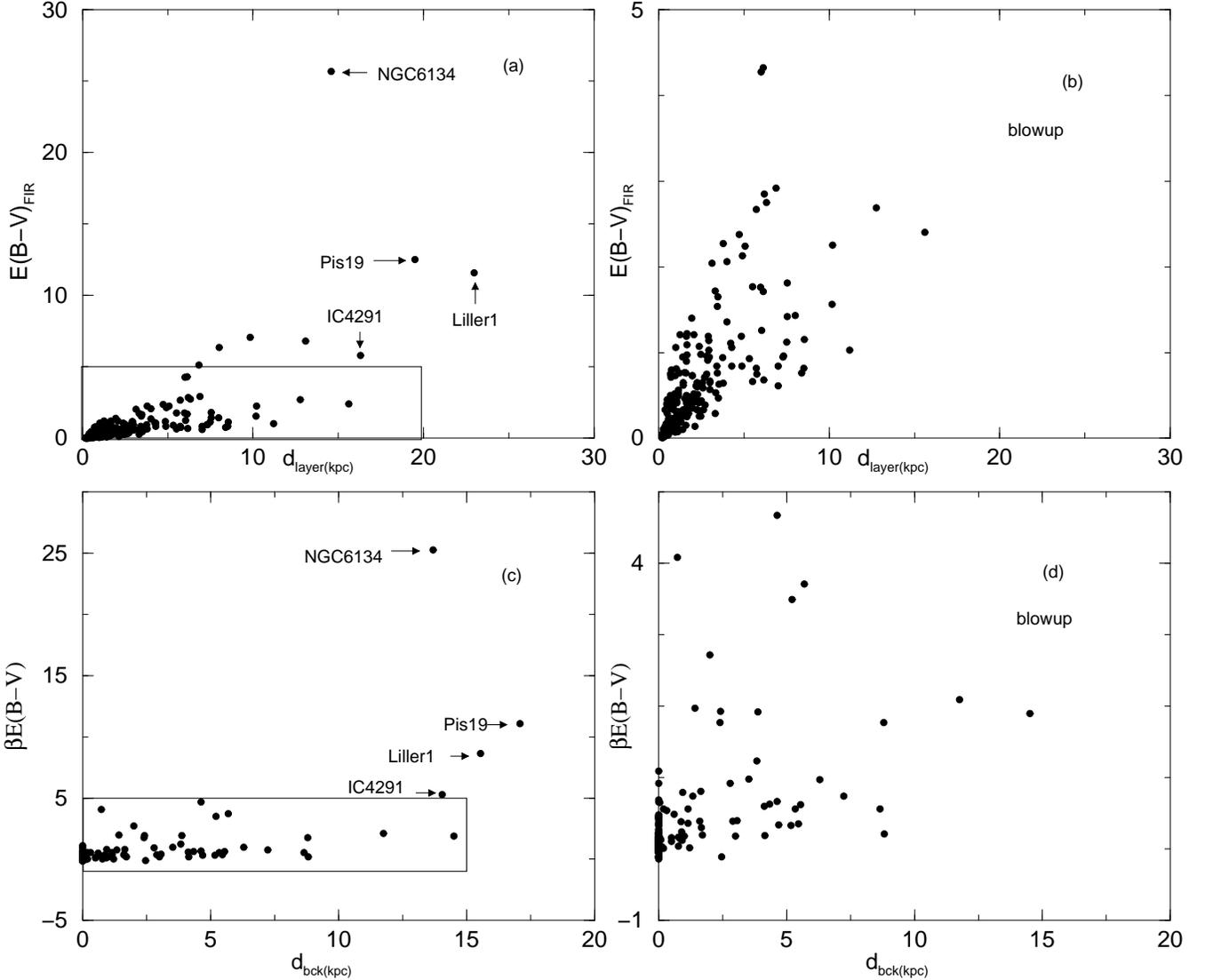}}
\caption[]{Panels: (a) Dust emission reddening E(B-V)$_{\rm FIR}$ as a function of the distance in the cluster line of sight from the true Plane up to the dust layer edge  d$_{\rm layer}$,  assuming a dust layer height of $|Z'|$ = 200 pc; (b) blowup of (a); (c) $\beta$E(B-V) as a function of background path between the cluster and the disk edge in that direction (d$_{\rm bck}$); (d) blowup of (c).}
\label{fig8}
\end{figure*}

\begin{table}
\caption[]{Distance from the plane and lines of sight for clusters with $\beta E(B-V) > 0.30$.}
\begin{scriptsize}
\label{tab4}
\renewcommand{\tabcolsep}{0.9mm}
\renewcommand{\arraystretch}{0.8}
\begin{tabular}{lccc}
\hline\hline
\\
Clusters within dust layer&&&\\
\\
\hline
\\
NAME&$|Z'|$&d$_{\rm layer}$&d$_{\rm bck}$\\
&(pc)&(kpc)&(kpc)\\
\\
\hline
Terzan\,4&    182 &   8.0 &   0.7 \\
Liller\,1$^*$&      6 &  22.9 &  15.5 \\
Terzan\,1&    106 &   9.8 &   4.6 \\
Terzan\,5&    120 &   6.0 &   2.4 \\
Terzan\,6&    188 &   5.7 &   0.3 \\
UKS\,1&    113 &  13.1 &   5.7 \\
Terzan\,9&    155 &   6.3 &   1.4 \\
Terzan\,10&    141 &   6.8 &   2.0 \\
NGC\,6544&     85 &   6.1 &   3.5 \\
Terzan\,12&    109 &   6.2 &   2.8 \\
Be\,81&     87 &   6.9 &   3.9 \\
IC\,4756&     55 &   1.6 &   1.2 \\
NGC\,6802&     30 &   6.1 &   5.2 \\
NGC\,6866&    193 &   1.5 &   0.1 \\
IC\,1369$^*$&      8 &  12.8 &  11.8 \\
NGC\,7226$^*$&     11 &  11.2 &   8.6 \\
Ki19$^*$&     18 &  10.2 &   8.8 \\
NGC\,7762&     93 &   1.6 &   0.9 \\
Cr\,463&     73 &   1.0 &   0.6 \\
IC\,166$^*$&      5 &   8.5 &   5.5 \\
Ki\,7$^*$&     24 &   7.5 &   5.3 \\
NGC\,1496$^*$&     18 &   7.5 &   6.3 \\
NGC\,2194&     93 &   5.7 &   3.1 \\
Tr\,5$^*$&     63 &   7.3 &   4.7 \\
NGC\,2236$^*$&     79 &   7.3 &   4.1 \\
NGC\,2112&    159 &   1.0 &   0.2 \\
Cr\,110&     52 &   7.5 &   5.5 \\
NGC\,2286&     72 &   6.0 &   3.8 \\
Haf\,8&     43 &   5.5 &   4.3 \\
Haf\,6$^*$&     28 &   8.4 &   5.2 \\
NGC\,2360$^*$&     17 &   8.5 &   7.2 \\
NGC\,2527&     35 &   3.5 &   2.9 \\
NGC\,2533&     50 &   6.2 &   4.6 \\
NGC\,2477&     99 &   2.3 &   1.1 \\
Pi\c{s}\,2&   150 &   3.8 &   0.9 \\
NGC\,2660&    136 &   4.2 &   1.3 \\
NGC\,3496$^*$&      7 &  15.6 &  14.5 \\
IC\,4291&     27 &  16.3 &  14.0 \\
Pi\c{s}\,19$^*$&      2 &  19.5 &  17.1 \\
NGC\,5823&     45 &   3.1 &   2.4 \\
NGC\,5822&     67 &   2.5 &   1.6 \\
NGC\,6005&    125 &   4.3 &   1.6 \\
NGC\,6134&     12 &  14.6 &  13.7 \\
\hline
\\
Clusters outside dust layer&&& \\
\\
\hline
Lyng\aa\,7&     311 &    4.3 &    0.0 \\
NGC\,6144&    2801 &    0.7 &    0.0 \\
NGC\,6256&     384 &    3.3 &    0.0 \\
NGC\,6355&     696 &    2.1 &    0.0 \\
Terzan\,2&     280 &    4.7 &    0.0 \\
HP\,1&     252 &    5.1 &    0.0 \\
NGC\,6380&     569 &    3.4 &    0.0 \\
Tonantzintla\,2&     367 &    3.5 &    0.0 \\
NGC\,6401&     847 &    2.8 &    0.0 \\
Palomar\,6&     214 &    6.0 &    0.0 \\
Djorgovski\,1&     227 &    4.9 &    0.0 \\
ESO456-SC38&     226 &    4.9 &    0.0 \\
NGC\,6553&     254 &    4.0 &    0.0 \\
NGC\,6749&     265 &    5.5 &    0.0 \\
Palomar\,10&     295 &    4.0 &    0.0 \\
Be\,54&     259 &    2.9 &    0.0 \\
NGC\,7044&     226 &    3.0 &    0.0 \\
\hline
\end{tabular}
\end{scriptsize}
\begin{list}{}
\item  Notes to Table 4: *lines of sight are truncated considering a galactic disk radius R = 15 kpc.
\end{list}
\end{table}

\subsection{Directions of some reddened young open clusters}

Since the young disk is considerably thinner than the old disk (Janes \& Phelps 1994, Friel 1995) it is worthwhile to study some interesting cases. We discuss some of the most reddened optical open clusters.

NGC\,3603 and Westerlund\,2 are clusters embedded in H\,II region complexes, where internal reddening is important. NGC\,3603 ($\ell$ = 291.61$^{\circ}$,  {\it b} = -0.52$^{\circ}$) is located at a distance d$_{\rm sun}$ = 7 kpc and has E(B-V) = 1.44 from the CMD (Melnick et al. 1989), which comprises both the internal and foreground reddening. From the integrated spectrum Santos \& Bica (1993) obtained a foreground reddening of E(B-V)$_{\rm f}$ = 1.18, implying an internal reddening E(B-V)$_{\rm i}$ = 0.26, by using a template spectrum which included internal absorption.
Westerlund\,2 ($\ell$ = 284.27$^{\circ}$, {\it  b} = -0.33$^{\circ}$) at a distance d$_{\rm sun}$ = 5.7 kpc has E(B-V) = 1.67 from the CMD (Moffat et al. 1991).  Piatti et al. 1998b derived  E(B-V)$_{\rm f}$ = 1.40 and E(B-V)$_{\rm i}$ = 0.27 by means of an integrated spectrum analysis.

Westerlund\,1 ($\ell$ = 339.55$^{\circ}$, {\it b} = -0.40$^{\circ}$) is possibly the most reddened open cluster which can be optically observed. By means of CMDs and integrated spectrum Piatti et al. (1998b) derived E(B-V) = 4.3 and d$_{\rm sun}$ = 1.0 kpc.

From Schlegel et al.'s (1998) reddening map we obtained very high reddening values derived for these young disk objects at very low galactic latitudes: E(B-V)$_{\rm FIR}$ = 59.7, 65.7 and 12.3 respectively for NGC\,3603, Westerlund\,2 and Westerlund\,1. The extremely high E(B-V)$_{\rm FIR}$ for NGC\,3603 and Westerlund\,2 are probably related to lines of sight intercepting dust cloud cores (Sect.2), presumably the molecular clouds from which they were formed. Since Westerlund\,1 is projected close to the Plane not far from the galactic center direction, its high E(B-V)$_{\rm FIR}$ can be explained by the dust cumulative effect produced by a series of spiral arms and the Molecular Ring in that direction (Sect.2.1).

\subsection{Reddening in the Sagittarius Dwarf direction}

The globular clusters associated with the Sagittarius Dwarf  are  indicated in 
Table 1. Their E(B-V) values derived from the stellar content are comparable
to those derived from the dust emission. Only a small systematic difference occurs, in the sense that  values  derived  from dust emission are larger by  $\beta$E(B-V) = 0.01-0.02. Assuming that M\,54, Terzan\,8, Arp\,2 and Terzan\,7 are slightly foreground or within   Sagittarius itself,  this sets 
a very low upper limit to the dust content in Sagittarius, in agreement with the fact that it is
very  depleted in H I (Koribalski et al. 1994).

\subsection{Evidence for dark clouds with $|Z'| > 200$ pc}

Assuming the Milky Way  dust layer height to be  $|Z'| = 200$ pc, Table 4 presents the 60 clusters with $\beta E(B-V) > 0.30$, showing their height from the Plane $|Z'|$,  the distance in the cluster line of sight from the true Plane up to the dust layer edge  d$_{\rm layer}$, and the path length behind the cluster up to the disk border d$_{\rm bck}$. They are separated in two groups, one formed by clusters within the dust layer (10 globular and 33 old open clusters) and the other by clusters outside it (15 globular and 2 old open clusters).  
As discussed in Sect.4.1 the large $\beta$E(B-V) values for clusters within the dust layer can be explained by dust clouds behind the clusters. The clusters outside the dust layer with large $\beta$E(B-V) values (see also their distribution in Fig.7 and the wing in the distribution above the dust layer) might be explained by higher $|Z'|$ dust clouds. A well-known example is  the high latitude dust cloud  Draco Nebula at a height from the Plane 300 to 400 pc (Gladders et al. 1998).
In addition, star forming complexes as traced by means of Wolf-Rayet stars indicate that they are concentrated within a height from the Plane of 225 pc, but some attain 300 pc (Conti \& Vacca 1990). 

 Assuming a dust layer height  $|Z'| = 300$ pc there would remain only 7 clusters outside the dust layer (Table 4). This corresponds to 3\% of the total sample of 250 star clusters in the present study. They are all globular clusters: Lyng\aa\,7, NGC\,6144, NGC\,6256, NGC\,6355, NGC\,6380, Tonantzintla\,2 and NGC\,6401. Among these clusters only NGC\,6144 is very far from the plane at $|Z'| \approx 2.8$ kpc (Table 4). Although NGC\,6144 has no CCD photometry yet, the photographic CMD  and the integrated light reddening estimates (Harris 1996 and references therein)  are consistent at about E(B-V) $\approx$ 0.32. We suspect that E(B-V)$_{\rm FIR}$ = 0.71 for this cluster is overestimated, arising from foreground dust heated above the upper limit 21 $^{\circ}$K (Schlegel et al. 1998) by the  hot star $\sigma$ Scorpii. Indeed, NGC\,6144 is seen through the edge of the $\rho$ Ophiuchi dark cloud complex in the association Upper Scorpius at distance d$_{\rm sun}$ = 125 pc (de Zeeuw et al. 1999). The cluster pathsight crosses the reflection nebula illuminated by the red supergiant Antares, also designated as IC\,4606, Cederblad\,132 or vdB-RN\,107 in the catalogue of reflection nebulae by van den Bergh (1966). The neighbouring star $\sigma$ Scorpii is double (B2\,III + O9.5\,V), and ionizes the H\,II region Sh\,2-9 (Sharpless 1959), or Gum\,65 (Gum 1955). This hot double star also has its reflection nebula component Cederblad\,130 (vdB-RN\,104), which almost overlaps with the Antares reflection nebula. It is possible that dust grains in the direction of NGC\,6144 are being heated by this particular configuration. At any rate CCD photometry of NGC\,6144 is necessary to definitely establish the reddening E(B-V) affecting the cluster stars.

 Howk \& Savage (1999) detected high-Z dust structures in a sample of 7 edge-on spiral galaxies (NGC\,891, NGC\,3628, NGC\,4013, NGC\,4217, NGC\,4302, NGC\,4565 and NGC\,4634). The thickness of the dust lanes in these galaxies is in the range $500 <  2 \times |Z| < 900$ pc. The high-Z dust features have typical dimensions of hundreds of parsecs and are located at heights in the range 500-1450 pc.  The present study indicates the need of a Milky Way dust layer of thickness $2 \times |Z| \approx 600$ pc in order to explain the lines of sight of 97\,\% of the known intermediate age and old clusters. The remaining 3\,\% would require some higher Z clouds.

\section{Concluding remarks}

Since the 100 $\mu$m dust emission reddening maps of Schlegel et al. (1998) provide  whole-sky reddening estimates with relatively high angular resolution it is important to explore them in detail for a better understanding of the dust properties in different directions.

We provided an overview of the distribution along the galactic plane and in some interesting latitude directions. The accumulation of dust clouds in different arms and large structures such as the Molecular Ring can be distinguished. Individual dust complexes, including their cores, have E(B-V)$_{\rm FIR}$ values compatible with those of embedded clusters derived from infrared photometry. An exception is the Nuclear Region where the temperature in the Central Molecular Zone appears to be underestimated in the Schlegel et al.'s temperature maps.     

The 100 $\mu$m dust emission reddening maps  provide E(B-V)$_{\rm FIR}$ values compatible to E(B-V) derived from the stellar content of globular and old open cluster for $\approx$ 75\,\% of the 250 directions probed in the present study ($\beta E(B-V) < 0.30$). The values for most high-latitude clusters ($|{\it b}|> 20^{\circ}$) are in good agreement; these are objects in general outside the disk dust layer, so that all dust in the line of sight is sensitive to both methods. An interesting exception is NGC\,1901 which is outside the dust layer and has a E(B-V)$_{\rm FIR}$ much larger than E(B-V). The background dust source in this case is the LMC disk.

The differences between the dust emission and stellar content reddening values occur most frequently for low latitude clusters ($|{\it b}|< 20^{\circ}$). 
Brightness selection effects due to reddening and distance particularly
 affect the open cluster sample. In the existing catalogues many
 intermediate age open clusters are yet to be studied, while future infrared surveys should reveal many new open clusters. These distant
 objects are expected to have increasing reddening values, but the $\beta$E(B-V) values should decrease since the pathsight behind the cluster within the dust layer decreases. Most of the known globular clusters have CMDs, but infrared surveys should reveal some new ones in disk and bulge zones. Those far in the disk should behave like the heavily reddened far open clusters.

Bandwidth effects on the intrinsic reddening law have implications on the
 results for larger extinctions. The fact that most of the points follow a 1:1
 FIR/Optical relation up to E(B-V) = 1.0 (Fig.5) suggests that this effect should become important for reddening values beyond this limit. An additional
 complication for objects so heavily reddened that the stellar content can be
 studied only in the infrared is that the transformations to optical
 reddening values depend on grain properties.

From the spatial distribution of available objects and their relative positions with respect to the dust layer we conclude that background dust clouds are probably responsible for these differences. A dust layer with thickness $2 \times |Z| \approx$ 600 pc is required to explain the distribution of $\approx$ 97\,\% of the sample. Some additional higher Z dust clouds, like the Draco Nebula, would also be required to explain the rest.

The present study of reddening in star cluster directions suggests that the Milky Way is similar in dust layer thickness and occurrence of some high-Z dust structures to edge-on spirals studied by Howk \& Savage (1999). In particular the Milky Way dust layer may be thicker than previously thought.

\begin{acknowledgements}
We acknowledge support from the Brazilian institution CNPq. We thank an anonymous referee for interesting remarks.
\end{acknowledgements}

%
%sssssssssssssssssssssssssssss REFERENCESsssssssssssssssssssssssssssssss
%

\end{document}